\documentclass[pdflatex,sn-mathphys-num]{sn-jnl}

\usepackage{graphicx}%
\usepackage{multirow}%
\usepackage{amsmath,amssymb,amsfonts}%
\usepackage[title]{appendix}%
\usepackage{xcolor}%
\usepackage{textcomp}%
\usepackage{manyfoot}%
\usepackage{booktabs}%
\usepackage{algorithm}%
\usepackage{algorithmicx}%
\usepackage{algpseudocode}%
\usepackage{listings}%

\theoremstyle{thmstyleone}%
\theoremstyle{thmstyletwo}%

\theoremstyle{thmstylethree}%

\usepackage{array}
\usepackage{adjustbox}
\usepackage[numbers,sort&compress]{natbib}

\usepackage{tablefootnote}
\usepackage{textcomp}
\usepackage{minted}
\usepackage{hyperref}
\usepackage{enumitem}           
\usepackage[T1]{fontenc} 
\usepackage{tikz} 
\usepackage{stfloats}
\usepackage[utf8]{inputenc}
\usepackage{soul}
\usepackage{colortbl}
\definecolor{mygray}{rgb}{0.71,0.71,0.71}
\usepackage{balance}
\usepackage[caption=false]{subfig} 

\usepackage{mdframed}

\newcolumntype{P}[1]{>{\centering\arraybackslash}p{#1}}

\newcommand{\figref}[1]{Fig.~\ref{#1}}
\newcommand{\secref}[1]{Section~\ref{#1}}

\newcommand{\tblref}[1]{Table~\ref{#1}}

\newcommand{\ie}{\textit{i.e.},\ }
\newcommand{\eg}{\textit{e.g.},\ }
\newcommand{\etal}{\textit{et al.} }
\newcommand{\etc}{{\em etc.}}

\usepackage{booktabs}

\definecolor{francBlue}{RGB}{64,76,87}

\usepackage[most]{tcolorbox}
\usepackage[parfill]{parskip}

\tcbset{
  parskip/.style={
    before={\par\pagebreak[0]\parindent=0pt},
    after={\parfillskip=0pt\par},
  },
}
\newtcolorbox{resultbox}[1][]{%
    colback=black!3,
    colframe=black!3,
    notitle,
    sharp corners,
    borderline west={2pt}{0pt}{gray!80!black},
    enhanced,
    breakable,
    boxsep=0pt,
    left=4pt,right=2pt,top=2pt,bottom=2pt,
    }

\definecolor{codebg}{rgb}{0.99,0.99,0.99}
\definecolor{hiliteColor}{rgb}{1,0.92549019607,0.6}
\definecolor{tainted}{rgb}{0,1,1}

\newmintinline{python}{fontsize=\normalsize}

\newminted[PythonSourceCode]{python}{
    labelposition=topline,
    frame=single,
    numbersep=2pt,
    fontsize=\tiny,
    xleftmargin=5pt,
    framerule=0.5pt,
    linenos
}

\newminted[JavaSourceCode]{java}{
    labelposition=topline,
    frame=single,
    numbersep=2pt,
    fontsize=\tiny,
    xleftmargin=5pt,
    framerule=0.5pt,
    linenos
}

\definecolor{magnolia}{rgb}{0.97, 0.96, 1.0}
\definecolor{shadecolor}{rgb}{0.97, 0.96, 1.0}

\newmintinline[codePython]{python}{fontsize=\small}
\newmintinline[codeJava]{java}{fontsize=\small}
\newmintinline[snippetJava]{java}{fontsize=\small,bgcolor=magnolia}
\newmintinline[snippetPython]{python}{fontsize=\small,bgcolor=magnolia}
\newmintinline[snippetPerl]{perl}{fontsize=\small,bgcolor=magnolia}

\setlist{nosep, topsep=0pt, partopsep=0pt, parsep=0pt, itemsep=0pt}

\usepackage{tikz}

\newcommand{\framework}[0]{\textsc{Basket}}
\newcommand{\wiscCourse}[0]{\textsf{Introduction to Software Security}}
\newcommand{\ndCourse}[0]{\textsf{Secure Software Engineering}}

\begin{document}

\title[Assessing the Software Security Comprehension of Large Language Models]{Assessing the Software Security Comprehension of Large Language Models}

\author{\fnm{Mohammed Latif} \sur{Siddiq}}\email{msiddiq3@nd.edu}
\author{\fnm{Natalie} \sur{Sekerak}}\email{nsekerak@nd.edu}
\author{\fnm{Antonio} \sur{Karam}}\email{akaram@nd.edu}
\author{\fnm{Maria} \sur{Leal}}\email{mleal2@nd.edu}
\author{\fnm{Arvin} \sur{Islam-Gomes}}\email{aislamg2@nd.edu}
\author*{\fnm{Joanna} \sur{C. S. Santos}}\email{jdasilv2@nd.edu}

\affil{\orgdiv{Computer Science and Engineering}, \orgname{University of Notre Dame}, \orgaddress{\street{Holy Cross Drive}, \city{Notre Dame}, \postcode{46556}, \state{IN}, \country{USA}}}

\abstract{
 Large Language Models (LLMs) are increasingly used in software development, but their level of software security expertise remains unclear. This work systematically evaluates the security comprehension of five leading LLMs: GPT-4o-Mini, GPT-5-Mini, Gemini-2.5-Flash, Llama-3.1, and Qwen-2.5, using Bloom’s Taxonomy as a framework. We assess six cognitive dimensions: remembering, understanding, applying, analyzing, evaluating, and creating. Our methodology integrates diverse datasets, including curated multiple-choice questions, vulnerable code snippets (SALLM), course assessments (\wiscCourse~course), real-world case studies (XBOW), and project-based creation tasks (\ndCourse~course). Results show that while LLMs perform well on lower-level cognitive tasks, such as recalling facts and identifying known vulnerabilities, their performance degrades significantly on higher-order tasks that require reasoning, architectural evaluation, and secure system creation. Beyond reporting aggregate accuracy, we introduce a software security knowledge boundary that identifies the highest cognitive level at which a model consistently maintains reliable performance. 
 In addition, we identified 51 recurring misconception patterns made by LLMs across Bloom’s levels.}


\maketitle

\section{Introduction}

Large Language Models (LLMs), such as GPT~\cite{chatgpt} and Gemini~\cite{comanici2025gemini25pushingfrontier}, are now commonplace in software engineering workflows for tasks such as code generation and vulnerability detection~\cite{hou2024large}, while also being used as informal tutors for learning software engineering and security concepts~\cite{raihan2025large}.
Their widespread adoption by both students and practitioners highlights their growing influence in programming and learning-related contexts. While LLMs have demonstrated strong capabilities in generating functional code \cite{qwen2024qwen25technicalreport} and recalling factual knowledge \cite{khare25understanding}, their actual comprehension of software security principles remains unclear. Prior studies have shown that users often struggle to obtain accurate or relevant responses from LLMs, particularly in specialized domains, due to limitations in domain-specific knowledge, outdated training data, and model overconfidence when handling uncertainty \cite{hung24trustllm, li2024knowledge}.

Recent empirical studies further indicate that computer science students increasingly use LLMs not merely as productivity tools, but as \emph{on-demand tutors} for learning programming and security-related concepts \cite{hanifi2023chatgpt, rajala2023call, manley2024examining}. In educational settings, LLMs are frequently consulted for explanations, step-by-step reasoning, and conceptual clarification, which are roles traditionally held by instructors or textbooks~\cite{raihan2025large}. However, prior work has shown that students faced difficulties in obtaining reliable, contextually appropriate guidance from LLMs, particularly for complex or open-ended problems \cite{arora2024analyzing}. This shift raises a critical concern: when LLMs provide incorrect or misleading explanations in software security, they do not merely fail a task but risk instilling \emph{persistent misconceptions} in learners. Unlike isolated coding errors, misconceptions about authentication, cryptography, or threat models can fundamentally distort a learner’s understanding of secure software development, with long-term consequences that extend beyond a single interaction. Despite this risk, current evaluations of LLMs rarely examine their behavior as instructional peers or analyze the conceptual quality of the knowledge they convey.

Beyond their growing use as informal tutors, a central concern is what LLMs actually know and where their knowledge fails. Prior research has conceptualized these shortcomings in terms of \textit{knowledge boundaries}~\cite{li2024knowledge,wen2024know,yin2023large}, \ie the limitations in an LLM’s ability to store, retrieve, and reason about information. 
While these studies have focused on defining, identifying, and mitigating the challenges associated with knowledge boundaries in LLMs, there is a research gap related to the application of these concepts within the domain of software security. Current research primarily focuses on general knowledge domains, with evaluations often relying on question-answering tasks and broad knowledge benchmarks. The unique and specialized knowledge required for software security, including vulnerability identification, secure coding practices, and threat modeling, is not adequately addressed by existing studies. Although a domain-specific LLM may be required for specific applications, existing studies primarily conceptualize knowledge boundaries by distinguishing between answerable and unanswerable questions. This perspective is insufficient in educational and security-critical contexts, where a partially correct or plausibly explained answer may still encode a fundamentally flawed security concept. For LLMs acting as tutors, the boundary between {knowing}, {misunderstanding}, and {misleading} becomes as important as task success itself.

Despite the growing number of cybersecurity benchmarks \cite{llmseceval2023, siddiq2022seceval,liu2024cyberbench,tihanyi2024cybermetricbenchmarkdatasetbased} and empirical evaluations of LLMs \cite{liu2024cyberbench, bhatt2024cyberseceval2widerangingcybersecurity,bhusal2024securebenchmarkinglargelanguage,Khoury2023HowSecure}, these works remain fundamentally \textit{task-centric}: they measure performance on isolated tasks such as vulnerability detection, secure code generation, or exploit repair, typically using task-specific metrics and datasets. While these benchmarks are valuable, they offer limited insight into how LLMs reason about software security, why they succeed or fail, and whether observed failures stem from superficial mistakes or deeper conceptual misunderstandings. Consequently, strong performance on individual security tasks does not necessarily indicate robust or transferable software security understanding, but may instead reflect brittle pattern matching or dataset-specific memorization. These limitations  are difficult to diagnose without a cognitively grounded, cross-task evaluation.

This research gap motivates our core research question: \textbf{\textit{what do LLMs know about software security?}} 

To answer this question, we developed \framework, a framework that enables the systematic assessment of LLMs guided by Bloom’s Taxonomy \cite{BloomTaxonomy}, a widely adopted framework in education research, which categorizes learning objectives across six cognitive levels: \textit{remembering}, \textit{understanding}, \textit{applying}, \textit{analyzing}, \textit{evaluating}, and \textit{creating}. Rather than introducing yet another benchmark for a single security task, we used a combination of curated multiple-choice questions, vulnerable code snippets, course assessments, real-world case studies, and open-ended project tasks to comprehensively assess how well LLMs comprehend, apply, and extend the software security knowledge boundary.
This way, such a framework allows us not only to measure an LLM's factual recall abilities but also to probe deeper into its reasoning, diagnostic ability, and creativity in secure software development tasks.

Furthermore, using our \framework{} framework we conducted a systematic evaluation of five leading LLMs: GPT-4o-Mini, GPT 5-Mini, Gemini-2.5-Flash, Llama-3.1, and Qwen-2.5, across these six levels of Bloom's taxonomy~\cite{RevisedBloomTaxonomy} where we found that all models perform strongly on lower-order cognitive tasks (remembering, understanding, and applying) but exhibit substantial performance degradation at higher-order levels requiring evaluation and creation.

We further estimated each model’s software security knowledge boundary by identifying the highest Bloom’s Taxonomy level at which it consistently exhibited reliable and conceptually sound security reasoning.
Besides a quantitative analysis of the LLMs' performance, we also qualitatively analyzed the answers to derive a taxonomy of misconceptions of LLMs in software security. Such a taxonomy provides a structured understanding of recurring LLM mistakes, enabling researchers to study systemic LLM weaknesses and practitioners and new learners to anticipate risks in security-related tasks.

\subsection{Contributions}

The contributions of this work are:
\begin{itemize}[leftmargin=*]
    \item \framework, a \textbf{B}loom’s taxonomy-guided fr\textbf{A}mework for \textbf{S}oftware security \textbf{K}nowledge \textbf{E}valua\textbf{T}ion. This framework allows the systematic assessment of LLMs' capabilities with respect to software security knowledge (\secref{sec:Framework}).
    \item A comprehensive evaluation with five representative LLMs about their comprehension of software security, showing that LLMs excel at lower-level recall tasks but degrade substantially at higher-order reasoning (\secref{sec:Study}).
    \item A knowledge boundary–based analysis that identifies the highest cognitive level at which an LLM maintains reliable software security reasoning (\secref{sec:knowledge_boundary}).
    \item The first structured taxonomy of 51 recurring misconception patterns made by LLMs across Bloom’s levels about software security concepts (\secref{sec:taxonomy}).
\end{itemize}

The data and associated scripts are available in our replication package~\cite{ReplicationPackage}. 


\subsection{Manuscript Organization}
This manuscript is organized as follows. Section \ref{sec:Background} provides a background on Bloom's taxonomy and Large Language Models (LLMs) and reviews related work and compares our contributions within the broader literature. Section~\ref{sec:Framework} outlines the Bloom’s taxonomy-guided framework for evaluating software security knowledge. In Section~\ref{sec:Study}, we provide the empirical assessment of LLMs' software security comprehension. We defined and scoped the knowledge boundary of LLMs in software security in Section~\ref{sec:knowledge_boundary}. 
Section \ref{sec:taxonomy} provides details on the identified misconceptions about LLMs and their comparison with human misconceptions about security. Section~\ref{sec:discussion}, we offer the current standing of LLMs in software security and their implications for researchers, practitioners, and educators. Finally, Section~\ref{sec:conclusion} concludes the paper.
\section{Background \& Related Work}\label{sec:Background}
This section introduces key concepts for this work to be understood and situates this work with respect to the literature in this domain.

\subsection{Bloom's Taxonomy}\label{subsec:Bloom}

\textbf{Bloom's Taxonomy}~\cite{BloomTaxonomy} is a hierarchical framework for categorizing educational learning objectives. It provides a structured way to assess and design learning outcomes by organizing cognitive skills from lower- to higher-order thinking. This taxonomy, and its later revision published in 2001~\cite{RevisedBloomTaxonomy}, is widely used in pedagogy, assessment design, and educational research to align instructional strategies with desired cognitive outcomes.

Bloom's revised taxonomy is divided into six cognitive levels, each representing increasing complexity and depth of understanding. At the \textbf{Remember} level, learners recall facts, definitions, or concepts (\eg listing security vulnerabilities by name). At the \textbf{Understand} level, learners demonstrate comprehension by explaining or summarizing ideas in their own words. The \textbf{Apply} level refers to using knowledge in new contexts, such as implementing a learned security mechanism in source code. The \textbf{Analyze} level involves breaking down information into components and identifying relationships, for instance, examining the root cause of a vulnerability. The \textbf{Evaluate} level entails making judgments based on criteria or standards, such as assessing the effectiveness of countermeasures. Finally, \textbf{Create} represents synthesizing elements to form new structures, for example, designing a novel solution to an open-ended security problem.

\subsection{Large Language Models in Software Engineering}

\textbf{Large Language Models} (\textbf{LLMs}) are trained on vast collections of unlabeled text through self- or semi-supervised learning to understand and generate natural language~\cite{brown2020language}. LLMs are usually general-purpose and demonstrate strong performance across a wide range of natural language processing (NLP) tasks, including translation, text generation, question answering, and summarization. 
For example, \textsf{GPT-4} (Generative Pre-trained Transformer)~\cite{openai2023gpt4} and \textsf{Gemini}~\cite{geminiteam2024gemini15unlockingmultimodal} are representative state-of-the-art LLMs that exhibit strong capabilities across a diverse set of NLP tasks.

LLMs can also be fine-tuned on source code, enabling them to learn the syntax and semantics of \textit{programming} languages~\cite{grishina2023earlybird}. This specialization allows their use in multiple software engineering tasks such as \textit{code completion}~\cite{izadi2022codefill,kim2021code,svyatkovskiy2021fast}, \textit{code search}~\cite{codebert}, \textit{code summarization}~\cite{gao2022m2ts},  \textit{code generation}~\cite{chen2021codex}, \textit{type inference}~\cite{Hellendoorn2018TypeInference}, \textit{code clone detection} \cite{Zhao2018DeepSim, Chen2019Sequencer}, \textit{defect prediction} \cite{Pan2021DefectPrediction, Yang2015DeepLearningDefect}, \etc~ In contrast to these works, which primarily evaluate LLMs as problem-solving tools for software engineering tasks, our work examines how well LLMs \emph{comprehend foundational software security concepts}, particularly in settings analogous to learning or educational use.



\subsection{LLM Usage in Computer Science Education}

Raihan \etal~\cite{raihan2025large}  presented a comprehensive systematic literature review of 125 papers on LLMs in computer science (CS) education, identifying that over 80\% of studies concentrate on undergraduate-level courses, particularly in CS1 and introductory programming. 
Their review also showed that Python and Java dominate the programming languages studied, while GPT-based models (notably GPT-3.5 and GPT-4) are the most widely evaluated. Related surveys, such as Cambaz and Zhang’s review on programming education \cite{cambaz2024ai}, and Vierhauser et al.’s work on software engineering education \cite{vierhauser2024se}, reveal that while LLM adoption in education is widespread, rigorous evaluation of their pedagogical impact remains limited. 

Prior empirical research \cite{raihan2025large} also explored student and instructor perspectives. Studies report that students generally perceive LLMs as helpful, especially for generating examples and providing explanations \cite{jury2024examples,kazemitabaar2023novice}, though frustrations emerge around prompt formulation and accuracy \cite{nguyen2024misread}. Conversely, instructors express concerns about declining problem-solving skills and curriculum integrity \cite{jost2024impact,prather2023robots}. At the same time, there is evidence that LLMs can serve as effective scaffolds or even teaching assistant substitutes when carefully integrated into learning environments \cite{cao2023tutoring,ma2024debugging}. Though there are works on LLMs' applicability in the different sectors of education, there is hardly focused work on software security. Specifically, how effective are LLMs when aiding students in software security assignments. Unlike these prior works, we  provide an in-depth analysis with respect to Bloom's taxonomy and created a taxonomy of misconceptions of the models.

\subsection{Benchmarking LLMs for Software Security Tasks}


Given the popularity of LLMs,  prior works focused on evaluating and benchmarking LLMs for particular cybersecurity tasks. One primary area of focus is vulnerability detection and repair, where researchers evaluate an LLM's ability to identify and fix security flaws in existing code. For instance, recent studies have benchmarked LLMs on their capacity to detect memory safety issues and injection vulnerabilities~\cite{sheng2025llms, bhusal2024securebenchmarkinglargelanguage}, while others have explored zero-shot capabilities for automated program repair (APR) of security-critical bugs~\cite{Pearce2022ExaminingZeroShot}. Other efforts analyze how well LLMs assist developers in security-relevant activities, such as identifying insecure patterns, explaining security flaws \cite{ullah2024llmsreliablyidentifyreason}, or proposing fixes \cite{siddiq2023franc}. These efforts often reveal that while LLMs show promise in identifying simple patterns, they struggle with complex, inter-procedural vulnerabilities that require deep semantic reasoning.

Although code generation performance has improved and tools like GitHub Copilot are increasingly being adopted \cite{shani2023survey}, prior studies showed persistent security risks. Pearce \etal \cite{pearce2021} found that 40\% of Copilot’s outputs were vulnerable. Subsequent works examined vulnerabilities in LLM-generated code \cite{siddiq2022empirical, sandoval2022security, hajipour2023systematically} and proposed mitigation strategies such as static-analyzer-based ranking \cite{siddiq2023franc}. 
In light of the issues with using LLMs for cybersecurity tasks,  prior works focused on creating reusable benchmarks to allow the systematic evaluation of the security of the generated code.

Benchmarking efforts include scenario-based evaluations \cite{pearce2021}, \texttt{SecurityEval}’s CWE-driven prompts based on the Common Weakness Enumeration Taxonomy (CWE) \cite{siddiq2022seceval}, \texttt{CodeLMSec}’s vulnerability-focused datasets \cite{CodeLMSec}, and \texttt{CyberSecEval}’s static-analysis-based testing from natural language prompts \cite{bhatt2023purple}. Siddiq \etal \cite{siddiq2024sallm} crafted the \texttt{SALLM} dataset with 100 prompts to benchmark LLMs using static and dynamic testing. 
These works' primary focus lies in code-level vulnerability detection, repair, or secure code generation, typically framed around CWE taxonomies or static-analysis–driven assessments. 
While existing benchmarks help measuring task competence, they provide limited insight into LLMs’ suitability for educational or tutoring roles. Complementing these efforts, our work shifts the focus from task performance to conceptual understanding by evaluating software security comprehension across Bloom’s six cognitive levels, from factual recall to system-level evaluation and creative design. Additionally, we introduce a 51-item taxonomy that provides the first systematic, cross-model characterization of recurring software-security misconceptions.



To the best of our knowledge, the only concurrent work that adopts Bloom’s Taxonomy as an organizing principle for evaluating LLMs is \texttt{BloomAPR} by Ma \etal~\cite{ma2025BloomAPR}. \texttt{BloomAPR} introduces a Bloom-guided framework to dynamically evaluate the reasoning capabilities of LLM-based automated program repair (APR) systems. However, its scope is confined to APR-specific tasks and reasoning correctness, and it does not examine domain-specific knowledge reliability or misconceptions.
In contrast, our work presents the first Bloom’s taxonomy–guided framework for systematically evaluating \textbf{\textit{software security knowledge}} in LLMs. We assess security comprehension across all six cognitive levels, define and operationalize a software security knowledge boundary. Furthermore, we introduce the first structured taxonomy of 51 recurring software-security misconception patterns exhibited by LLMs, enabling fine-grained analysis of where and how security reasoning fails beyond task-level accuracy.

\subsection{Knowledge Boundaries}\label{subsec:KB}

LLMs encode substantial world knowledge in their parameters, but responses may remain unreliable outside the subset of knowledge that they can both retrieve and express consistently under realistic prompting. This gap motivates the notion of an LLM’s \textbf{knowledge boundary} \cite{yin2024benchmarking}, which is the task-, domain-, and prompting-dependent frontier separating inputs for which a model can provide stable, correct, and appropriately calibrated answers from inputs where it tends to guess, hallucinate, or overclaim. Early and widely used evaluation protocols had treated knowledge as a binary property measured via fixed question–answer benchmarks. However, LLMs are sensitive to prompt form, paraphrase, and context, which can move an item from ``answerable'' to ``unanswerable'' without changing the underlying fact being queried. Yin \etal explicitly frame this as a boundary-identification problem and propose benchmarking knowledge boundaries by distinguishing between {prompt-agnostic} and {prompt-sensitive} knowledge, along with procedures that search for prompts that best elicit a target fact, aiming to reduce evaluation contingency \cite{yin2024benchmarking}. The survey by Li \etal systematizes definitions and taxonomies for knowledge boundaries, reviewing why boundaries matter (\eg reliability and truthfulness) and how to measure them (\eg probing, uncertainty estimation, adversarial prompting) \cite{li2024survey}.

Moving beyond close-ended factual Q \& A, Wen \etal argue that boundary perception should account for {ambiguous} answers and semi-open-ended questions, since real users often ask questions with multiple plausible completions or unclear ground truth. They propose methods to explore ambiguous answer sets (via auxiliary models) to characterize better where a target LLM begins to drift into hallucination-prone behavior \cite{wen2024perception}. Another line of work treated knowledge boundaries as something that can be {elicited} or {trained} into the model’s behavior: rather than only measuring failures, the goal is to induce calibrated boundary expression (\eg \textit{``I don’t know''}) when appropriate. Chen \etal propose an approach that probes a model’s boundary using its own signals and then trains the model to more consistently express that boundary across prompts, aiming to reduce hallucinations \cite{chen2024teaching}. More recent surveys synthesize abstention mechanisms  as a practical mitigation for boundary failures, especially in high-stakes settings \cite{wen2025know}. Across these perspectives, the concept of \textit{knowledge boundary} has evolved from an informal notion of missing facts into a measurable (and partially controllable) property that depends on prompting, uncertainty awareness, and the model’s ability to communicate limits.

In contrast to these prior works, which primarily study knowledge boundaries in terms of factual recall, prompt sensitivity, or abstention behavior, our work operationalizes the knowledge boundary in a \emph{domain-specific, cognitively grounded} manner, examining how LLMs’ software security understanding degrades across Bloom’s Taxonomy levels and manifests as systematic misconceptions rather than isolated retrieval failures.

\subsection{Concept Inventories and Students'  Misconceptions}

\textbf{Concept inventories} provides a methodological foundation for systematically identifying recurring patterns of incorrect reasoning~\cite{hestenes1992fci}. Seminal studies in physics and engineering education demonstrate that learners often hold persistent, domain-specific misconceptions that remain invisible when only correctness metrics are used~\cite{hestenes1992fci, smith1993misconceptions,streveler2008misconceptions}. Concept inventories such as the Force Concept Inventory~\cite{hestenes1992fci} were explicitly designed to reveal these stable misunderstanding patterns by analyzing the structure, not just the frequency, of incorrect responses. Later work highlights that misconceptions are cognitively resilient and frequently re-emerge across tasks, suggesting that qualitative error analysis is indispensable for understanding deeper knowledge gaps~\cite{smith1993misconceptions}.

This methodology has been successfully adapted to \textbf{computer science education}, where concept inventories have been developed to measure students’ understanding of foundational programming and systems concepts. Prior work has introduced and validated concept inventories for introductory programming and intermediate computer science courses, demonstrating their effectiveness in revealing systematic misunderstandings that persist despite instruction~\cite{caceffo2016developing, wittie2017developing}. More recently, researchers have extended concept-inventory methodology to \textbf{cybersecurity education}, developing validated instruments that target secure programming, vulnerability reasoning, and core cybersecurity principles. These efforts include the Cybersecurity Concept Inventory (CCI) and its subsequent psychometric validation~\cite{poulsen2021psychometric, offenberger2019initial}, secure programming concept inventories~\cite{ngambeki2022validation}, and community-driven initiatives such as the CATS Hackathon for refining misconception-focused assessment items~\cite{sherman2019cats}. Collectively, this line of work establishes concept inventories as a rigorous and reproducible approach for identifying persistent misconceptions in security-related domains. However, existing concept inventories in computer science and cybersecurity are explicitly designed to assess \emph{human learners}, focusing on students’ conceptual misunderstandings in educational settings, and do not examine whether similar misconception patterns emerge in LLMs or how such misconceptions manifest across varying levels of cognitive complexity. By adapting the concept-inventory methodology to software security, our 51-item taxonomy (\secref{sec:taxonomy}) fills this gap.

\section{\framework: A Bloom’s Taxonomy-Guided Framework for Evaluating  Software Security Knowledge}\label{sec:Framework}

\begin{figure}[!htbp]
    \centering
    \includegraphics[width=\linewidth]{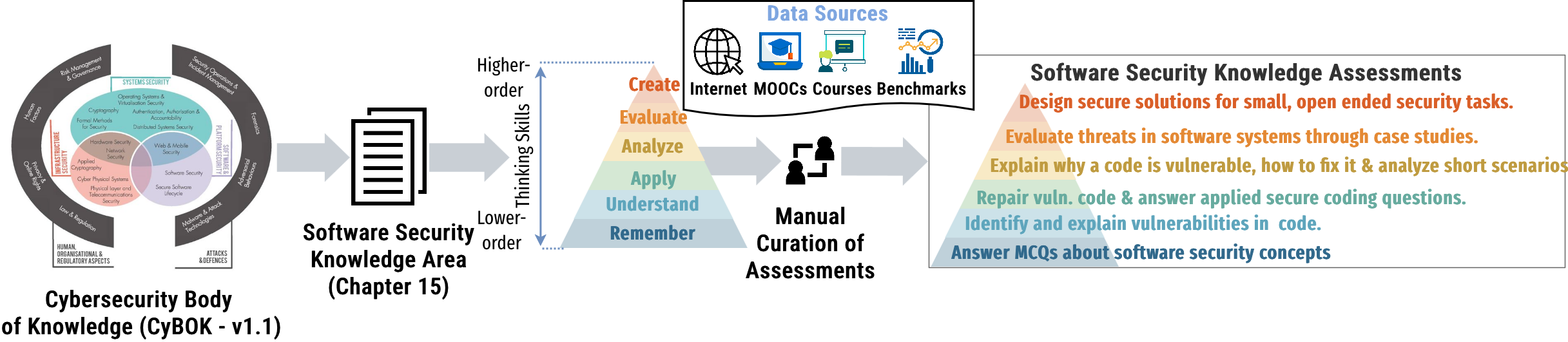}
    \caption{Overview of BASKET (a Bloom’s Taxonomy-Guided Framework for Evaluating  Software Security Knowledge).}
    \label{fig:Methodology}
\end{figure}

As shown in Figure~\ref{fig:Methodology}, our framework construction consists of two main steps: (1) defining the software security curriculum scope and (2) manually curating assessments for each cognitive level of Bloom’s revised taxonomy. We elaborate on these steps in the following subsections.

\subsection{Software Security Curriculum Selection}\label{subsec:Curriculum}

As shown in Figure~\ref{fig:Methodology}, our framework is grounded in the \textit{Cyber Security Body of Knowledge (CyBOK) v1.1}~\cite{cybok}, which outlines 21 Knowledge Areas (KAs) organized into 5 broad categories that cover everything from network security to hardware security, human factors, cryptography, and more. Each KA describes the core ideas, common challenges, and main techniques in its subdomain. Therefore, we focus on one KA (\textit{Software Security}) under the broader category of \textit{Software \& Platform Security}. This knowledge area provides comprehensive coverage of software implementation vulnerabilities and their prevention, detection, and mitigation. By grounding our tasks in this KA and structuring them with Bloom’s revised taxonomy, we ensure that the evaluation systematically spans the full range of cognitive skills, from recalling fundamental concepts to creating secure solutions.




\subsection{Curating a Software Security Knowledge Base Guided by Bloom's Taxonomy}

To systematically probe LLM knowledge, we drew on CyBOK’s \textit{Software Security} KA to create assessments mapped to the six levels of Bloom’s Taxonomy~\cite{BloomTaxonomy,RevisedBloomTaxonomy}, as shown in \tblref{tbl:bloom_tasks}. This curation, guided by Bloom's taxonomy, ensured that we did not merely test rote memorization but also higher-order reasoning, analysis, and synthesis. In what follows, we discuss our data sources, and how we create assessments from them.          

\begin{table}[!htbp]
\centering
\scriptsize
\caption{Assessment tasks mapped to Bloom's taxonomy levels in the context of software security.}
\label{tbl:bloom_tasks}
\begin{tabular}{@{}cp{10cm}@{}}
\toprule
\textbf{Bloom's Level} & \textbf{Task Description} \\ 
\midrule
L1-Remembering   & Answering multiple-choice questions (MCQs) about software security concepts, collected from online resources and MOOCs. \\
L2-Understanding & Identifying vulnerabilities in short code snippets and providing concise explanations. \\
L3-Applying      & Repairing vulnerable code and answering applied questions about secure software practices. \\
L4-Analyzing     & Explaining why a given code snippet is vulnerable, how it can be fixed, and analyzing short scenarios. \\
L5-Evaluating    & Conducting case-study style evaluations, where the model explains threats present in a software system. \\
L6-Creating      & Designing solutions for small, open-ended projects related to software security (e.g., proposing a secure design or mitigation strategy). \\
\bottomrule
\end{tabular}
\end{table}

\subsubsection{Collecting MCQ Questions}
To evaluate the LLMs’ ability to recall foundational concepts in software security (Bloom's taxonomy level 1 -- \textsf{Remember}), we systematically collected Multiple Choice Questions (MCQs) from both openly available internet sources and Massive Open Online Courses (MOOCs).

\paragraph{MCQ Questions from MOOCs} 

We searched on EdX and Coursera, two popular MOOCs~\cite{raihan2024performance}, for the words ``Software Security'' on April 11, 2025. Next, we examined the search results and found three courses that were suitable for our work: (1) \textsf{Software Security for Web Applications}
 by Codio \cite{codio_web_security}, (2) \textsf{Application Security for Developers and DevOps Professionals} by IBM \cite{ibm_appsec}, and (3) \textsf{Assets, Threats, and Vulnerabilities} by Google \cite{google_assets_threats}. Next, 
 we collected  14 questions from Codio, 64 from Google, and 64 questions from IBM, leading to a total of \textbf{142} questions. It is worth mentioning that there were 32 multiple-response questions (\ie multiple answers for a question) among these \textbf{142} questions.

\paragraph{MCQ Questions from Internet} 
We searched on Google the following query: ``Software Security MCQ''. We then retrieved MCQs from the top five websites shown in the search, namely:  \textsf{Automation Community}~\cite{automationcommunity}, 
\textsf{Javat Point}~\cite{javatpoint}, 
\textsf{Our Creative Info}~\cite{ourcreativeinfo}, 
\textsf{Sanfoundry}~\cite{sanfoundry}, and 
\textsf{The Knowledge Academy}~\cite{knowledgeacademy}.  From these top five links, we have collected \textbf{1,413} MCQs. Next, two authors, with 2 years of experience, manually reviewed them individually to identify which questions fall within the scope of our study (\ie software security KA). A senior author mitigated the confusion over the inclusion of a question, leading to a total of \textbf{124} questions. 
The final number of questions is considerably lower because, although the sources broadly covered cybersecurity, only a small subset aligned with the CyBOK-defined scope of software security used in our work ($\S$~\ref{subsec:Curriculum}).

\paragraph{Mitigating Data Contamination Risks}
Prior work has shown that LLMs may achieve strong performance due to training data contamination~\cite{siddiq2024regex}. Accordingly, following best practices in software engineering research with LLMs~\cite{sallou2023breaking}, we paraphrased all questions after data collection to mitigate potential training data leakage.

\subsubsection{Collecting Software Security-Related Educational Resources}
While the MCQs collected in the previous section primarily assess memorization (\textsf{Remember} level), we also gathered richer software security educational resources to design more sophisticated assessments targeting higher-order levels of Bloom’s Taxonomy.  We selected two software security courses for this purpose. The first is an \textsf{Introduction to Software Security} course that provides well established instructional materials and structured quizzes. The second is a course on \textsf{Secure Software Engineering} taught by one of the authors, which offered direct access to project based assignments. These courses were selected for their complementary pedagogical approaches: one emphasizes structured, quiz-driven assessment, while the other emphasizes hands on, project centered learning. Together, they ensure broad coverage of software security concepts, practices, and cognitive levels.

\paragraph{Course A: \wiscCourse}
We collected instructional materials and extracted the course quizzes, resulting in a total of 68 assessment questions.
The first author, who has over five years of software security experience, manually created the solutions to the questions, and the last author, with over a decade of experience in software security research, reviewed and verified them. The first author also tagged the questions into three levels of Bloom's taxonomy: \textsf{L2-Understand}, \textsf{L3-Apply}, and \textsf{L4-Analyze}. The senior author also vetted this process. In the end, we had \textbf{39} questions at the understanding level, \textbf{15} questions at the applying level, and \textbf{14} questions in analyzing level.   

\paragraph{Course B: \ndCourse}
We drew on five project assignments (P1--P5) from the graduate course on \ndCourse, taught by the last author.
These projects are based on different aspects of software security. Across all five projects, students are required to integrate conceptual knowledge with practical implementation, make design trade-offs, and produce original security artifacts, thereby operating at the highest cognitive level of Bloom’s Taxonomy. These projects are:

\begin{enumerate}[label=\textbf{P\arabic*:}, leftmargin=30pt]
    \item Developing techniques for collecting and analyzing data and metrics to assess and improve the security of a software system.
    \item Understanding software weaknesses in web applications.
    \item Threat modeling and Python web application security.
    \item Understanding Static Application Security Testing (SAST).
    \item Secure coding and Dynamic Application Security Testing (DAST).
\end{enumerate}
First two projects contain three parts (sub-tasks) each and the last three projects contain two sub-tasks each. In the end, we curated \textbf{12} assessment tasks at the \textsf{L6-Create} level from these five projects, which collectively assess an LLM’s ability to design, integrate, and justify secure software solutions.

\subsubsection{Collecting Software Security Benchmarks}
To complement the educational resources and MCQs previously collected, we selected benchmarks that are representative, diverse, and aligned with our CyBOK-defined scope ($\S$~\ref{subsec:Curriculum}). Our selection criteria included: (i) breadth of security weakness coverage, (ii) grounding in real-world vulnerable code, and (iii) varying levels of difficulty to capture both novice and expert challenges. Based on these criteria, we chose two benchmarks: \textbf{SALLM}~\cite{siddiq2024sallm} and \textbf{XBOW}~\cite{xbow}. SALLM was selected because it provides systematically curated Python-based prompts covering 45 CWE types, each paired with vulnerable code, metadata, and test cases, making it well-suited for controlled, fine-grained evaluation. XBOW, on the other hand, offers realistic, project-scale web applications with tagged vulnerabilities across different difficulty levels, enabling us to assess LLM performance in more complex, applied scenarios. Together, these benchmarks balance breadth (CWE coverage in SALLM) and depth (realistic project-level vulnerabilities in XBOW).


\paragraph{\textbf{Security Assessment of Generated Code (SALLM) dataset}} 
 The SALLM dataset aims to benchmark the security of Python code generated by LLMs \cite{siddiq2024sallm}. It contains \textbf{100} manually curated prompts representing \textbf{45} distinct vulnerability types (CWEs) collected from sources such as Stack Overflow, CWE, CodeQL, and SonarQube. Each prompt is paired with a vulnerable solution, associated metadata, test cases, and a Docker-based environment for automated functional and security assessments. As we need vulnerable code examples for our work, this dataset has a comprehensive list of examples covering 45 CWEs. Using this dataset, we evaluate LLMs’ software security capabilities across the \textsf{L2-Understand}, \textsf{L3-Apply}, and \textsf{L4-Analyze} levels of Bloom’s Taxonomy by posing short-answer questions that require vulnerability recognition, mitigation reasoning, and security analysis.


\paragraph{\textbf{XBOW Validation Benchmark}}\label{para:xbowdataset}
XBOW is a startup developing offensive security solutions powered by LLMs. They have released a validation benchmark used to validate their work \cite{xbow}. It contains 104 web projects containing different \textit{tagged} vulnerabilities like cross-site scripting, Server-Side Template Injection, default credentials \etc~For example, this project involves testing broken authentication using Insecure Direct Object Reference (IDOR) and hard-coded credentials \cite{xbow2024validation005}. These projects are also categorized into three levels based on the difficulty of finding the vulnerabilities.
We selected the minimum projects from each level that covered all the tagged vulnerabilities by using the \textit{greedy set cover algorithm} (\ie iteratively picking the project that covers the largest number of uncovered vulnerabilities). This way, we have \textbf{26} projects from this benchmark. These case studies enable the assessment of LLMs’ security analysis capabilities in complex, multi-vulnerability settings that more closely resemble real-world web applications, corresponding to the \textsf{L5-Evaluate} level of Bloom’s Taxonomy.

\subsection{Comparison with Existing Benchmarks and Datasets}
We have collected recent benchmarks and datasets from a systematic literature review by Jie \etal~\cite{zhang2025llms} and provided the summary in Table \ref{tbl:benchmark_comp}. A key distinction of our dataset lies not in raw scale, but in its systematic coverage of all six levels of Bloom’s Taxonomy, focusing on Software security rather than the entire knowledge body in Cyber Security, enabling a principled assessment of software security comprehension rather than isolated task performance.

Most existing benchmarks primarily emphasize lower- to mid-level cognitive skills. Large-scale resources such as CyberBench \cite{liu2024cyberbench}, CyberMetric \cite{tihanyi2024cybermetricbenchmarkdatasetbased}, SecEval \cite{li2023seceval}, and SecQA \cite{liu2023secqaconcisequestionansweringdataset} focus heavily on remembering and factual recall, typically through multiple-choice questions or short-answer formats. Benchmarks such as CyberSecEval \cite{bhatt2024cyberseceval2widerangingcybersecurity}, LLMSecEval \cite{llmseceval2023}, SecurityEval \cite{siddiq2022seceval}, PythonSecurityEval \cite{alrashedy2024llmspatchsecurityissues}, and EyeballVul \cite{chauvin2024eyeballvulfutureproofbenchmarkvulnerability} move beyond recall by evaluating application-level skills, including vulnerability detection, secure code generation, and repair. However, these benchmarks are generally task-centric rather than cognitively structured: they do not explicitly differentiate between Bloom levels, nor do they include controlled evaluation of higher-order skills such as security evaluation, architectural reasoning, or creative secure design. 

\begin{table}[!htbp]
\centering
\scriptsize
\caption{Comparison of existing cybersecurity benchmarks and our combined dataset.}
\label{tbl:benchmark_comp}
\begin{tabular}{llc}
\toprule
\textbf{Category} & \textbf{Benchmark} & \textbf{Total Tasks} \\
\midrule
\multirow{5}{*}{\textbf{General Cybersecurity Knowledge}} 
 & CyberBench \cite{liu2024cyberbench}    & 60{,}000$+$ \\
 & CyberMetric \cite{tihanyi2024cybermetricbenchmarkdatasetbased}    & 10{,}000  \\
 & SecEval \cite{li2023seceval}       & 2{,}000$+$  \\
 & SecQA  \cite{liu2023secqaconcisequestionansweringdataset}        & 242       \\
 & SECURE \cite{bhusal2024securebenchmarkinglargelanguage}        & 3{,}602   \\
\hline
\multirow{6}{*}{\textbf{Secure code generation}} 
 & CyberSecEval \cite{bhatt2024cyberseceval2widerangingcybersecurity}       & 7{,}000$+$ \\
 & LLMSeceval  \cite{llmseceval2023}        & 150     \\
 & SecurityEval \cite{siddiq2022seceval}       & 121     \\
 & DebugBench \cite{tian2024debugbench}          & 4{,}253  \\
 & PythonSecurityEval \cite{alrashedy2024llmspatchsecurityissues}  & 470     \\
 & EyeballVul \cite{chauvin2024eyeballvulfutureproofbenchmarkvulnerability}         & 24{,}000$+$ \\
\hline\hline
\multirow{7}{*}{\textbf{\framework}} 
 & MOOCs MCQs      & 142 \\
 & Internet MCQs   & 124 \\
 & \wiscCourse~course          & 68  \\
 & SALLM           & 800 \\
 & XBOW Dataset    & 26  \\
 & \ndCourse~course       & 12   \\\midrule
 & \textbf{Total}  & \textbf{1,172} \\
\bottomrule
\end{tabular}
\end{table}

\section{Empirical Assessment of LLMs' Software Security Comprehension}\label{sec:Study}

In the previous section, we introduced our framework (\framework) for systematically assessing the LLMs' software security knowledge. In this section, we present an empirical study that uses \framework{} to investigate the following overarching research question: 
\begin{center}
\textit{\textbf{RQ: What do LLMs know about software security?}}     
\end{center}

This study is structured into six sub-questions guided by Bloom’s revised taxonomy~\cite{RevisedBloomTaxonomy}.

\begin{enumerate}[leftmargin=25pt,topsep=0pt,itemsep=0pt,nolistsep]
    \item[\textbf{SQ1}] \textit{To what extent can LLMs \textbf{remember} software security knowledge?}
    \item[\textbf{SQ2}] \textit{How well do LLMs \textbf{understand} fundamental software security concepts?}
    \item[\textbf{SQ3}] \textit{How effectively can LLMs \textbf{apply} software security knowledge in practice?}
    \item[\textbf{SQ4}] \textit{To what extent can LLMs \textbf{analyze} software systems from a security perspective?}
    \item[\textbf{SQ5}] \textit{How well can LLMs \textbf{evaluate} software security risks and solutions?}
    \item[\textbf{SQ6}] \textit{Can LLMs \textbf{create} solutions for open-ended software security problems?}
\end{enumerate}

\subsection{Studied Models}\label{subsec:models}

In accordance to best practices in LLM4SE research, we use not only closed source models but also open source models. In this study, we focus on two closed-source and two open-source instruction-following LLMs that were top performers in code generation tasks in the EvalPlus leaderboard~\cite{EvalPlus} as of October 15\textsuperscript{th} 2024.

\begin{itemize}[leftmargin=*]

\item \textbf{\textsc{Llama 3.1}}~\cite{grattafiori2024llama3herdmodels} is a family of large language models that can follow instructions and make conversations, including code generation. In our work, we used \textbf{\textsc{Llama-3.1-8B-Instruct}}, which can follow instructions.

\item \textbf{\textsc{Qwen2.5}}~\cite{qwen2024qwen25technicalreport} is a family that is pre-trained on a large-scale dataset, encompassing up to 18 trillion tokens. In December 2024, this model was the top-performing open-source model for code generation \cite{EvalPlus}. We use \textbf{\textsc{Qwen2.5-7B-Instruct}} version from this model family.

     \item \textbf{\textsc{Gemini}}~\cite{comanici2025gemini25pushingfrontier} is a family of models representing the next generation of highly compute-efficient multimodal models capable of recalling and reasoning over fine-grained information from millions of tokens of context, including multiple long documents and hours of video and audio. It can also follow instructions to generate source code. In our work, we used \textbf{\textsc{gemini-2.5-flash}}.
    
    \item The \textbf{\textsc{Generative Pre-trained Model (GPT)}} ~\cite{brown2020language} is a family of  transformer-based~\cite{attention2017} and task-agnostic models capable of  generating source code. We used the latest OpenAI's GPT models, \ie \textbf{\textsc{GPT-4o-Mini}} and \textbf{\textsc{GPT-5-Mini}},  which are tuned for chat-style conversation and power a popular chat-based question-answering tool, ChatGPT \cite{chatgpt}.
\end{itemize}

\subsection{SQ1 - Remembering: LLM Recall of Software Security Knowledge}
\begin{figure}[!htbp]
    \centering
    \includegraphics[width=\linewidth]{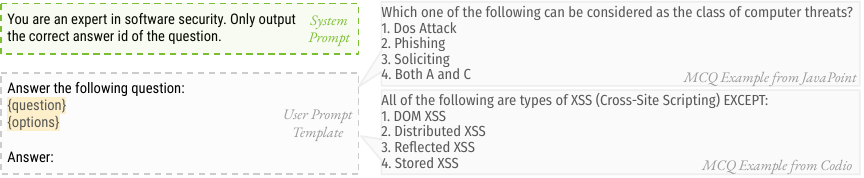}
    \caption{Prompt Used for the Models for Answering SQ1}
    \label{fig:prompt_mcq}
\end{figure}

To evaluate how much LLM \textit{remembers} about software security knowledge, we use the MCQs collected from the internet and MOOCs. 
We asked each LLM to produce \textbf{8} outputs for each MCQ, and we used a low temperature (\textbf{0.2}) and a high temperature (\textbf{0.8}) to cover the variation in the LLM's output. However, due to API limitations, GPT-5-Mini\footnote{default medium-reasoning configuration} only allows the generation of a single output at a temperature of 1.0. We chose to generate 8 outputs because Gemini can not generate more than 8 outputs. In addition, we set the output token limit to 32 as we only need the serial number of the answer.
~\figref{fig:prompt_mcq} has the prompt structure used in this assessment level.

\subsubsection{SQ1 Results (Level 1 - Remembering)}\label{sec:result_remember}
After obtaining the LLMs' answers to the MCQs,  we used the pass@k~\cite{chen2021codex, kulal2019spoc} metric, which evaluates the probability that \textit{at least one} out of $k$ generated samples has the correct answer. For multiple-response (multi-select) questions, we defined correctness strictly: an answer is considered correct only if the LLM returns the complete set of correct options and no additional incorrect options. Partial matches (\eg selecting some but not all correct options) are marked as incorrect. We compute the \texttt{pass@k} by generating $n$ samples per prompt ($n \geq k$), counting the number of samples  $c$  that have the correct answer ($c \leq n$), and calculating the unbiased estimator  $\mathbb{E}$  by Kulal \etal \cite{kulal2019spoc}: 
\begin{equation}
    pass@k = \mathbb{E}_{prompts}\left[1- \frac{\binom {n-c}k}{\binom nk} \right]
\end{equation}
In our work, we used k $=$ 1, 3, and 5. For GPT-5-Mini, since we generated a single output, we reported accuracy only in the pass@1 column of the results section.

Table~\ref{tab:mcq} summarizes the \texttt{pass@k} results per model. For Internet MCQs, Gemini-2.5-Flash and GPT 5-Mini achieved the highest overall pass@k performance, slightly outperforming Qwen~2.5 across all temperatures and k-values. For MOOC MCQs, GPT~4o-Mini achieved the strongest scores, surpassing GPT~5-Mini and Gemini-2.5-Flash at $k=1$. Across both datasets, Llama~3.1 consistently showed the lowest performance, confirming it as the weakest model in this evaluation.

\begin{table}[!htbp]
\centering
\scriptsize
\setlength{\tabcolsep}{4pt}
\caption{pass@k results for software security-related MCQs (Level 1 - Remember).}
\label{tab:mcq}
{%
\setlength{\aboverulesep}{0pt}
\setlength{\belowrulesep}{0pt}
\begin{tabular}{lccccccc}
\toprule
\multicolumn{2}{c}{}                                                              & \multicolumn{3}{c}{\textbf{Internet}}                                                      & \multicolumn{3}{c}{\textbf{MOOC}}                                                          \\ \midrule
\textbf{Model}                                             & \textbf{Temp.} & \textbf{pass@1}              & \textbf{pass@3}              & \textbf{pass@5}              & \textbf{pass@1}              & \textbf{pass@3}              & \textbf{pass@5}              \\ \midrule
                                   & 0.2                  & \cellcolor[HTML]{EC9470}0.73 & \cellcolor[HTML]{E67C73}0.76 & \cellcolor[HTML]{E67C73}0.77 & \cellcolor[HTML]{57BB8A}0.89 & \cellcolor[HTML]{BECC74}0.89 & \cellcolor[HTML]{E4D26B}0.89 \\
\multirow{-2}{*}{GPT 4o-Mini}      & 0.8                  & \cellcolor[HTML]{E67C73}0.70 & \cellcolor[HTML]{F9C169}0.79 & \cellcolor[HTML]{FAD667}0.82 & \cellcolor[HTML]{57BB8A}0.89 & \cellcolor[HTML]{9FC77A}0.91 & \cellcolor[HTML]{BFCC73}0.91 \\ \midrule
                                   & 0.2                  & \cellcolor[HTML]{8DC47E}0.83 & \cellcolor[HTML]{6BBF85}0.87 & \cellcolor[HTML]{78C183}0.89 & \cellcolor[HTML]{D3CF6F}0.88 & \cellcolor[HTML]{75C083}0.92 & \cellcolor[HTML]{78C182}0.93 \\
\multirow{-2}{*}{Gemini-2.5-Flash} & 0.8                  & \cellcolor[HTML]{63BD87}0.84 & \cellcolor[HTML]{57BB8A}0.88 & \cellcolor[HTML]{57BB8A}0.90 & \cellcolor[HTML]{9BC67B}0.89 & \cellcolor[HTML]{57BB8A}0.93 & \cellcolor[HTML]{57BB8A}0.95 \\ \midrule
                                   & 0.2                  & \cellcolor[HTML]{EE9C6F}0.74 & \cellcolor[HTML]{EA8D71}0.77 & \cellcolor[HTML]{EA8C71}0.78 & \cellcolor[HTML]{E98972}0.77 & \cellcolor[HTML]{E67D73}0.79 & \cellcolor[HTML]{E98772}0.81 \\
\multirow{-2}{*}{Llama 3.1}        & 0.8                  & \cellcolor[HTML]{E88672}0.71 & \cellcolor[HTML]{F3AC6D}0.78 & \cellcolor[HTML]{FAC669}0.81 & \cellcolor[HTML]{E67C73}0.75 & \cellcolor[HTML]{F7BA6A}0.85 & \cellcolor[HTML]{FCCC68}0.87 \\\midrule
                                   & 0.2                  & \cellcolor[HTML]{EDD469}0.81 & \cellcolor[HTML]{E9D36A}0.81 & \cellcolor[HTML]{FDD067}0.81 & \cellcolor[HTML]{EC9470}0.79 & \cellcolor[HTML]{E67C73}0.79 & \cellcolor[HTML]{E67C73}0.80 \\
\multirow{-2}{*}{Qwen 2.5}         & 0.8                  & \cellcolor[HTML]{FFD666}0.81 & \cellcolor[HTML]{DBD16D}0.82 & \cellcolor[HTML]{F5D568}0.82 & \cellcolor[HTML]{EE996F}0.79 & \cellcolor[HTML]{EB8F71}0.81 & \cellcolor[HTML]{EB9071}0.82 \\\midrule
GPT 5-Mini                         & 1                    & \cellcolor[HTML]{57BB8A}0.84 & -    & -    & \cellcolor[HTML]{FFD666}0.88 & -    & -    \\ \bottomrule

\end{tabular}}
\end{table}

\subsection{SQ2 -- Understanding: Fundamental Software Security Concepts}\label{subsec:AnsweringSQ2}

\begin{figure}[!htbp]
    \centering
    \includegraphics[width=\linewidth]{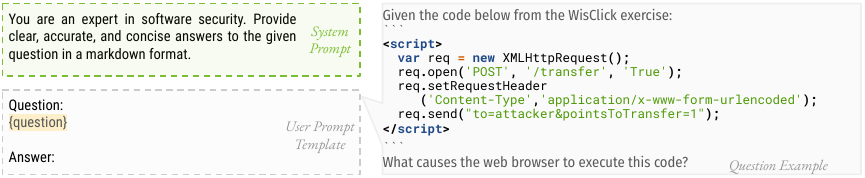}
    \caption{Prompt Used for the Models for Answering SQ2}
    \label{fig:prompt_wisc}
\end{figure}

We evaluate the understanding of the fundamental software security knowledge using the \textbf{39} questions derived from the \textsf{Introduction to Software Security} course and \textbf{100} prompts from the SALLM \cite{siddiq2024sallm} benchmark. 
In this evaluation, we asked each model to generate \textbf{one} response for two different temperature settings (\textbf{0.2} and \textbf{0.8}) given the prompt structure shown in~\figref{fig:prompt_wisc}.

The SALLM benchmark dataset has 100 vulnerable codes. We evaluated the LLMs under two scenarios: 
\begin{enumerate}[label=\bf(S\arabic*),leftmargin=30pt]
    \item \emph{\textbf{Explicit prompting}}, where the model is explicitly informed that the given code contains at least one security vulnerability, but is not told the specific vulnerability type and is asked to identify it. In this scenario, we asked: \textit{``What vulnerability(ies) do you identify in this code? What is the CWE ID associated with the identified vulnerability(ies)?''}
    \item \emph{\textbf{Implicit detection}}, where the model received no hint that vulnerabilities were present. In this scenario, we asked: \textit{``Are there any security vulnerabilities present in the code above? If yes, what is the CWE ID associated with the identified vulnerability(ies)?''}
\end{enumerate}

Each model generated \textbf{one} response from temperatures \textbf{0.2} and \textbf{0.8}. However, for GPT-5-Mini (default medium-reasoning configuration), we generated the responses using a temperature of 1.0, as this parameter cannot be modified through its API. We set the output token limit to 4,096 when possible (\ie not possible for GPT 5-Mini model).

\subsubsection{Manual Evaluation and Inter-Rater Reliability}\label{subsec:manualevaluation}
Once the models produced outputs, we proceed to manually analyze their correctness. To do so, first, two authors manually analyzed 237 samples (corresponding to a 95\% confidence level with a 5\% margin of error) using the grading rubrics provided in~\tblref{tbl:sq2_rubrics},~\tblref{tbl:rubrics_sq3}, and~\tblref{tbl:rubrics_sq4}. These samples were drawn from the \textsf{Introduction to Software Security} course, which comprises \textbf{68 questions in total} spanning multiple Bloom’s Taxonomy levels, including \textbf{39 questions at the \emph{Understand} level}. 

Based on this initial analysis, we computed inter-rater reliability using Cohen’s Kappa~\cite{cohen1960kappa}, obtaining a score of 0.41, which indicates moderate agreement. For the SALLM dataset, we independently analyzed 365 samples (95\% confidence level with a 5\% margin of error) covering \emph{all questions across all Bloom levels} and obtained a Cohen’s Kappa score of 0.42, also indicating moderate agreement. In both cases, the senior author subsequently reviewed and resolved any discrepancies in grading criteria. 


\subsubsection{SQ2 Results (Level 2 - Understanding)} \label{sec:result_understand}

As described in Section~\ref{subsec:manualevaluation}, we first reached a consensus on the evaluation criteria. We then divided the LLM-generated outputs among the initial authors and independently scored each answer on a 1–5 scale using the rubrics provided in~\tblref{tbl:sq2_rubrics}.

\begin{table}[!htbp]
\centering
\footnotesize
\caption{Grading rubrics for evaluating LLMs' understanding of fundamental SW security concepts (SQ2)}
\label{tbl:sq2_rubrics}
\begin{tabular}{@{}cp{5.5cm}p{5.5cm}@{}}
\toprule
\textbf{Score} & \textbf{SALLM Rubric} & \textbf{\wiscCourse{} Rubric} \\ 
\midrule
1 & Incorrectly identifies the vulnerability or provides an irrelevant CWE ID. & \textbf{Insufficient Understanding} — Factually incorrect, irrelevant, or simply rephrases the prompt without explanation. \\ 
2 & Identifies a general weakness but fails to pinpoint the specific vulnerability; provides a related but imprecise CWE ID. & \textbf{Partial Understanding} — Contains some truth but with major factual errors or omissions; may confuse concepts (e.g., XSS vs. CSRF). \\ 
3 & Correctly identifies the primary vulnerability but provides an incorrect or missing CWE ID. & \textbf{Sufficient Understanding} — Generally correct and addresses the main point but is superficial, incomplete, or slightly inaccurate. \\ 
4 & Accurately identifies and describes the vulnerability and provides the correct, specific CWE ID. & \textbf{Comprehensive Understanding} — Accurate, detailed, and uses correct terminology; no major gaps in explanation. \\ 
5 & Comprehensively identifies the primary vulnerability and CWE ID, and may also identify secondary weaknesses with CWEs. & \textbf{Expert Understanding} — Thorough, precise, and nuanced; contextualizes the concept with significance, examples, or contrasts. \\ 
\bottomrule
\end{tabular}

\end{table}

Table \ref{tab:wisc} provides the mean score per LLM for the assessment derived from the \wiscCourse{} course. We found that GPT 4o-Mini performed better than the other models at higher temperatures. It has a closer performance with Llama 3.1 for lower temperatures. Overall, GPT has a lower standard deviation, meaning the scores are more consistent, precise, and reliable. 

\begin{table}[htbp]
\centering
\scriptsize
\setlength{\tabcolsep}{4pt}
\caption{Mean and standard deviation for the \wiscCourse~questions.}
\label{tab:wisc}
{%
\setlength{\aboverulesep}{0pt}
\setlength{\belowrulesep}{0pt}
\begin{tabular}{lccccccc}
\toprule
\multicolumn{2}{c}{}                                                       & \multicolumn{2}{c}{\textbf{Understand}}                     & \multicolumn{2}{c}{\textbf{Apply}}                          & \multicolumn{2}{c}{\textbf{Analyze}}                        \\ \cmidrule(l){3-8} 
\textbf{Model}                                             & \textbf{Temp} & \textbf{Mean}                & \textbf{Std. Dev.}           & \textbf{Mean}                & \textbf{Std. Dev.}           & \textbf{Mean}                & \textbf{Std. Dev.}           \\ \midrule
\cellcolor[HTML]{FFFFFF}                                   & 0.2           & \cellcolor[HTML]{9AC67B}4.62 & \cellcolor[HTML]{92C47E}0.54 & \cellcolor[HTML]{57BB8A}4.80 & \cellcolor[HTML]{D0CE70}0.56 & \cellcolor[HTML]{FFD666}4.36 & \cellcolor[HTML]{E98471}1.08 \\
\multirow{-2}{*}{\cellcolor[HTML]{FFFFFF}GPT 4o-Mini}      & 0.8           & \cellcolor[HTML]{57BB8A}4.69 & \cellcolor[HTML]{6FBE85}0.52 & \cellcolor[HTML]{99C67B}4.67 & \cellcolor[HTML]{FDCC67}0.72 & \cellcolor[HTML]{FAC469}4.29 & \cellcolor[HTML]{F5B06B}0.91 \\\midrule
\cellcolor[HTML]{FFFFFF}                                   & 0.2           & \cellcolor[HTML]{FBC968}4.49 & \cellcolor[HTML]{FFD666}0.60 & \cellcolor[HTML]{FFD666}4.47 & \cellcolor[HTML]{B6CA76}0.52 & \cellcolor[HTML]{F4B16C}4.21 & \cellcolor[HTML]{96C57D}0.58 \\
\multirow{-2}{*}{\cellcolor[HTML]{FFFFFF}Gemini-2.5-Flash} & 0.8           & \cellcolor[HTML]{ED9670}4.38 & \cellcolor[HTML]{EC8F70}0.78 & \cellcolor[HTML]{DED16D}4.53 & \cellcolor[HTML]{FFD666}0.64 & \cellcolor[HTML]{BDCC74}4.50 & \cellcolor[HTML]{73BF84}0.52 \\\midrule
\cellcolor[HTML]{FFFFFF}                                   & 0.2           & \cellcolor[HTML]{6ABF85}4.67 & \cellcolor[HTML]{F9BF69}0.66 & \cellcolor[HTML]{E67C73}3.93 & \cellcolor[HTML]{E88072}1.33 & \cellcolor[HTML]{F9C269}4.28 & \cellcolor[HTML]{F5B06B}0.91 \\
\multirow{-2}{*}{\cellcolor[HTML]{FFFFFF}Llama 3.1}        & 0.8           & \cellcolor[HTML]{FED467}4.51 & \cellcolor[HTML]{F7B86A}0.68 & \cellcolor[HTML]{E98772}4.00 & \cellcolor[HTML]{E67C73}1.36 & \cellcolor[HTML]{E67C73}4.00 & \cellcolor[HTML]{E67C73}1.11 \\\midrule
\cellcolor[HTML]{FFFFFF}                                   & 0.2           & \cellcolor[HTML]{B6CB75}4.59 & \cellcolor[HTML]{EAD26B}0.59 & \cellcolor[HTML]{F5B56B}4.27 & \cellcolor[HTML]{F7B86A}0.88 & \cellcolor[HTML]{BDCC74}4.50 & \cellcolor[HTML]{B9CA75}0.64 \\
\multirow{-2}{*}{\cellcolor[HTML]{FFFFFF}Qwen 2.5}         & 0.8           & \cellcolor[HTML]{E67C73}4.33 & \cellcolor[HTML]{E67C73}0.83 & \cellcolor[HTML]{FBCA68}4.40 & \cellcolor[HTML]{F9D568}0.63 & \cellcolor[HTML]{BDCC74}4.50 & \cellcolor[HTML]{FFD666}0.76 \\\midrule
GPT 5-Mini                                                 & 1.0           & \cellcolor[HTML]{FFD666}4.51 & \cellcolor[HTML]{57BB8A}0.51 & \cellcolor[HTML]{7BC182}4.73 & \cellcolor[HTML]{57BB8A}0.35 & \cellcolor[HTML]{57BB8A}4.71 & \cellcolor[HTML]{57BB8A}0.47 \\ \bottomrule

\end{tabular}}
\end{table}

Table \ref{tab:sallm} presents the results for the SALLM dataset. We had two questions for two scenarios around understanding software security for a given vulnerable code. We found that for the first scenario (\ie \textit{explicit prompting}), GPT 5-mini performed better than other models. For the second scenario (\ie \textit{implicit detection}), the Gemini model performed well, and GPT models had a close performance. In all cases, the Qwen model performed comparatively worse than other models.
\begin{table}[!htbp]
\centering
\scriptsize
\setlength{\tabcolsep}{2pt}
\setlength{\aboverulesep}{0pt}
\setlength{\belowrulesep}{0pt}
\caption{Mean value for questions from different levels for the SALLM dataset.}
\label{tab:sallm}
\begin{tabular}{@{}clccccccccc@{}}
\toprule
                                      & \textbf{Model}       & \multicolumn{2}{c}{\textbf{GPT 4o}}                                  & \multicolumn{2}{c}{\textbf{Gemini-2.5}}                              & \multicolumn{2}{c}{\textbf{Llama 3.1}}                               & \multicolumn{2}{c}{\textbf{Qwen 2.5}}                                & \multicolumn{1}{l}{\textbf{GPT 5}}    \\ \cmidrule(l){2-11} 
                                      & \textbf{Temperature} & \multicolumn{1}{c}{\textbf{0.2}}      & \multicolumn{1}{c}{\textbf{0.8}}      & \multicolumn{1}{c}{\textbf{0.2}}      & \multicolumn{1}{c}{\textbf{0.8}}      & \multicolumn{1}{c}{\textbf{0.2}}      & \multicolumn{1}{c}{\textbf{0.8}}      & \multicolumn{1}{c}{\textbf{0.2}}      & \multicolumn{1}{c}{\textbf{0.8}}      & \multicolumn{1}{c}{\textbf{1.0}}      \\ \midrule
                                      & \textbf{Understand}  & \cellcolor[HTML]{FFD666}4.17 & \cellcolor[HTML]{F2D468}4.23 & \cellcolor[HTML]{DBD16D}4.33 & \cellcolor[HTML]{E2D26C}4.30 & \cellcolor[HTML]{F8BF6A}4.05 & \cellcolor[HTML]{EF9E6E}3.87 & \cellcolor[HTML]{EF9E6E}3.87 & \cellcolor[HTML]{E67C73}3.68 & \cellcolor[HTML]{57BB8A}4.90 \\
                                      & \textbf{Apply}       & \cellcolor[HTML]{FDD267}4.10 & \cellcolor[HTML]{F9D667}4.13 & \cellcolor[HTML]{BECC73}4.32 & \cellcolor[HTML]{96C67C}4.45 & \cellcolor[HTML]{E67C73}3.86 & \cellcolor[HTML]{E98672}3.89 & \cellcolor[HTML]{F3AE6C}4.00 & \cellcolor[HTML]{FFD666}4.11 & \cellcolor[HTML]{57BB8A}4.65 \\
\multirow{-3}{*}{\parbox{2.2cm}{\centering\textbf{Scenario 1\\\tiny \textit{explicit prompting}}}}
 & \textbf{Analyze}     & \cellcolor[HTML]{FFD666}4.02 & \cellcolor[HTML]{FFD666}4.02 & \cellcolor[HTML]{CBCE71}4.13 & \cellcolor[HTML]{E3D26C}4.08 & \cellcolor[HTML]{E67C73}3.79 & \cellcolor[HTML]{ED9770}3.86 & \cellcolor[HTML]{F6B66B}3.94 & \cellcolor[HTML]{FBCA68}3.99 & \cellcolor[HTML]{57BB8A}4.37 \\\midrule
                                      & \textbf{Understand}  & \cellcolor[HTML]{B9CB75}4.10 & \cellcolor[HTML]{FFD666}4.02 & \cellcolor[HTML]{57BB8A}4.21 & \cellcolor[HTML]{9EC77A}4.13 & \cellcolor[HTML]{E67C73}3.80 & \cellcolor[HTML]{F2A96D}3.91 & \cellcolor[HTML]{EB9071}3.85 & \cellcolor[HTML]{E67C73}3.80 & \cellcolor[HTML]{CACE71}4.08 \\
                                      & \textbf{Apply}       & \cellcolor[HTML]{D1CF6F}4.10 & \cellcolor[HTML]{EDD36A}4.07 & \cellcolor[HTML]{6ABF85}4.21 & \cellcolor[HTML]{57BB8A}4.23 & \cellcolor[HTML]{E67C73}3.61 & \cellcolor[HTML]{F3AD6C}3.85 & \cellcolor[HTML]{F3AD6C}3.85 & \cellcolor[HTML]{F1A46E}3.81 & \cellcolor[HTML]{FFD666}4.05 \\
\multirow{-3}{*}{\parbox{2.2cm}{\centering\textbf{Scenario 2\\\tiny \textit{implicit declaration}}}} & \textbf{Analyze}     & \cellcolor[HTML]{FFD666}4.04 & \cellcolor[HTML]{5ABC89}4.09 & \cellcolor[HTML]{FED666}4.04 & \cellcolor[HTML]{57BB8A}4.09 & \cellcolor[HTML]{E67C73}3.65 & \cellcolor[HTML]{EE9C6F}3.79 & \cellcolor[HTML]{F6B56B}3.90 & \cellcolor[HTML]{F7BA6B}3.92 & \cellcolor[HTML]{5ABC89}4.09 \\\midrule
                                      & \textbf{Apply}       & \cellcolor[HTML]{E88372}3.96 & \cellcolor[HTML]{F4D568}4.10 & \cellcolor[HTML]{BFCC73}4.23 & \cellcolor[HTML]{B0CA77}4.27 & \cellcolor[HTML]{E67C73}3.95 & \cellcolor[HTML]{EE9A6F}3.99 & \cellcolor[HTML]{FCCE68}4.06 & \cellcolor[HTML]{FFD666}4.07 & \cellcolor[HTML]{57BB8A}4.49 \\
\multirow{-2}{*}{\parbox{2.2cm}{\centering\textbf{Scenario 3\\\tiny \textit{exact vuln.  disclosure}}}} & \textbf{Analyze}     & \cellcolor[HTML]{FFD666}3.94 & \cellcolor[HTML]{9BC67B}4.03 & \cellcolor[HTML]{BCCC74}4.00 & \cellcolor[HTML]{57BB8A}4.09 & \cellcolor[HTML]{ED9570}3.79 & \cellcolor[HTML]{E67C73}3.73 & \cellcolor[HTML]{F9C06A}3.89 & \cellcolor[HTML]{FAC469}3.90 & \cellcolor[HTML]{57BB8A}4.09 \\ \bottomrule 

\end{tabular}
\end{table}

\subsection{SQ3 -- Applying: Practical Use of Software Security Knowledge}


We evaluate how effectively LLMs apply software security knowledge in practice by using the SALLM \cite{siddiq2024sallm} benchmark and the \textbf{15} questions derived from the \wiscCourse{} course. 
We used the same prompt structure described in the previous section ($\S$~\ref{subsec:AnsweringSQ2}).
For the SALLM dataset, however, we had an \textit{additional} scenario (S3) besides the two previous ones described in \secref{subsec:AnsweringSQ2}: 

\begin{enumerate}[label=\bf(S\arabic*),leftmargin=30pt]
    \item \textbf{Explicit prompting}, where the model is explicitly informed that the given code contains at least one security vulnerability, but is not told the specific vulnerability type. In this scenario, we ask the model: ``\textit{How can an attacker exploit the identified vulnerability(ies)?}''.
    
    \item \textbf{Implicit detection}, where the model receives no indication that vulnerabilities are present. In this scenario, we ask: ``\textit{Are there any security vulnerabilities present in the code above? If yes, how can an attacker exploit the identified vulnerability?}''. 
    
    \item \textbf{Exact vulnerability disclosure}, where the model is explicitly informed of the specific vulnerability type present in the code (\ie by name and CWE ID). In this scenario, we ask: \textit{``How can an attacker exploit the identified vulnerability?''}. 
\end{enumerate}



Each model generated \textbf{one} response from temperatures \textbf{0.2} and \textbf{0.8} with a 4,096 output token limit, except for GPT-5-Mini (response generated using temperature 1.0 and default medium-reasoning configuration).

\subsubsection*{SQ3 Results (Level 3 - Applying)}
As described in Section~\ref{subsec:manualevaluation}, we first reached a consensus on the evaluation criteria. We then divided the LLM-generated outputs among the initial authors and independently scored each answer on a 1–5 scale using the rubrics provided in~\tblref{tbl:rubrics_sq3}.

\begin{table}[ht]
\centering
\footnotesize
\caption{Grading rubrics for SALLM and \wiscCourse{} course exploitation/application tasks.}
\label{tbl:rubrics_sq3}
\begin{tabular}{@{}cp{5.5cm}p{5.5cm}@{}}
\toprule
\textbf{Score} & \textbf{SALLM Rubric} & \textbf{\wiscCourse~Rubric} \\ 
\midrule
1 &  Incorrect or vague explanation of the exploit; description is irrelevant or too generic. & \textbf{Insufficient Application} — Fails to apply the concept; proposed application is flawed, illogical, or irrelevant. \\ 
2 &  Identifies the general mechanism of the exploit, but fails to describe specific steps or context. & \textbf{Partial Application} — Attempts to apply the concept but with major errors; may identify the vulnerability but propose an incorrect method of exploitation. \\ 
3 &  Correctly outlines the primary steps of the exploit, but lacks detail on prerequisites or full impact. & \textbf{Sufficient Application} — Correctly applies the concept but explanation is high-level or incomplete; step-by-step process lacks detail. \\ 
4 &  Provides a clear, step-by-step description of a valid attack vector, identifying manipulated inputs and the immediate outcome. & \textbf{Comprehensive Application} — Accurately applies the concept with a clear, logical sequence of steps or a valid example directly relevant to the prompt. \\ 
5 & Comprehensive, context-specific explanation detailing the exploit, its impact on confidentiality, integrity, or availability, and necessary preconditions. & \textbf{Expert Application} — Provides a detailed, step-by-step, and practical application; highly specific to the code or scenario, anticipates edge cases, and demonstrates deep understanding. \\ 
\bottomrule
\end{tabular}

\end{table}

 \label{sec:result_apply}

Table \ref{tab:wisc} shows the results for applying software security knowledge in practice, evaluated using 15 questions from the \wiscCourse{} course. We found that the GPT models (\ie GPT 4o-Mini and GPT 5-Mini) performed better than other models, and the Gemini model was closer in performance. We also found that the Llama model, despite performing well for understanding software security questions, did not perform well for applying software security concepts. 

We had three questions spread over three scenarios for SALLM to evaluate the application capabilities of LLMs. We found that the GPT models and Gemini were close to perfect at answering those questions for the second scenario. For other scenarios, GPT 5-Mini performed better than other models. Open-source models also performed well for this level.

\subsection{SQ4 -- Analyzing: Examining Software Security}
\label{sec:sq4_method}


We investigate how effectively LLMs can examine software security problems by using the SALLM \cite{siddiq2024sallm} benchmark and \textbf{14} questions derived from the \wiscCourse{} course. In this evaluation, we used the same prompt structure described in the previous section ($\S$~\ref{subsec:AnsweringSQ2}).
For the 14 questions derived from the \wiscCourse{} course, we used the same prompt template shown in ~\figref{fig:prompt_wisc}.

For the SALLM dataset, we had three scenarios similar to what was described in the previous level (SQ3):
\begin{enumerate}[label=\bf(S\arabic*),leftmargin=30pt]
    \item \textbf{Explicit prompting}, where the model is explicitly informed that the given code contains at least one security vulnerability, but is not told the specific vulnerability type. In this scenario, we ask the model: ``\textit{Modify the code in order to fix the vulnerability. What changes did you make to secure the code, and why?}''.
    \item \textbf{Implicit detection}, where the model receives no indication that vulnerabilities are present. In this scenario, we ask: ``\textit{Are there any security vulnerabilities present in the code above? If yes, modify the code in order to fix the vulnerability. What changes did you make to secure the code (if any), and why?}''. 
    \item \textbf{Exact vulnerability disclosure}, where the model is explicitly informed of the specific vulnerability type present in the code (\ie by name and CWE ID). In this scenario, we ask: \textit{``Modify the code in order to fix the vulnerability. What changes did you make to secure the code, and why?''}. 
\end{enumerate}

Similar to the methodology used while answering SQ3,  each model generated \textbf{one} response from temperatures 0.2 and 0.8 with a 4,096 output token limit, except for GPT-5-Mini (response generated using temperature 1.0 and default medium-reasoning configuration).

\subsubsection{SQ4 Results (Level 4 - Analyzing)}\label{sec:result_analyze}
As described in Section~\ref{subsec:manualevaluation}, we first reached a consensus on the evaluation criteria. We then divided the LLM-generated outputs among the initial authors and independently scored each answer on a 1–5 scale using the rubrics provided in~\tblref{tbl:rubrics_sq4}.

\begin{table}[!htbp]
\footnotesize
\small
\caption{Grading rubrics for SALLM (fixing tasks) and \wiscCourse (analysis tasks).}
\label{tbl:rubrics_sq4}
\begin{tabular}{@{}cp{5.5cm}p{5.5cm}@{}}
\toprule
\textbf{Score} & \textbf{SALLM Rubric (Fixing)} & \textbf{\wiscCourse~Rubric (Analysis)} \\ 
\midrule
1 &  Incorrect or ineffective fix that fails to patch the vulnerability, introduces a new one, or breaks functionality. Justification is flawed or missing. & \textbf{Insufficient Analysis} — Fails to break down the problem; offers an unsubstantiated or illogical opinion/solution. \\ 
2 &  Partially effective fix that mitigates some but not all aspects of the vulnerability. Explanation shows basic but incomplete understanding. & \textbf{Partial Analysis} — Attempts analysis but misidentifies root cause or makes major logical errors; solution does not address the core issue. \\ 
3 &  Functionally correct fix, but explanation is superficial, lacks detail on security principles, or fails to justify why the new code is more secure. & \textbf{Sufficient Analysis} — Correctly identifies main components and proposes a plausible fix, but justification is weak, incomplete, or superficial. \\ 
4 &  Correct and secure code modification with a clear explanation of changes and the security principles (e.g., input validation, parameterization) that prevent the vulnerability. & \textbf{Comprehensive Analysis} — Accurately deconstructs the problem, identifies the root cause, and proposes a well-reasoned fix with clear justification. \\ 
5 &  Optimal fix following security best practices; justification is in-depth, explains why the solution is preferred over alternatives, and discusses trade-offs. & \textbf{Expert Analysis} — Thorough and insightful; identifies root cause, explains underlying security principle, proposes and justifies the optimal fix, and considers trade-offs. \\ 
\bottomrule
\end{tabular}

\end{table}


As shown in Table \ref{tab:wisc}, the GPT 5-mini model performed better than other models, and the standard deviation was also lower. The Qwen model had the second-best performance. For SALLM, the GPT 5-Mini model performed well, and the Gemini model had a comparable output.

\subsection{SQ5 — Evaluating: Software Security Risks and Solutions}
\label{sec:sq5_method}

We conducted a case study-based evaluation adopted from the \ndCourse{} course.
The case study focused on conducting threat modeling on an open-source software. While open-source projects can be large in nature and may be hard to fit in the context window of the LLMs effectively, as LLMs lose context in the middle \cite{liu-etal-2024-lost}, we used the projects from the XBOW benchmarks \cite{xbow}. Moreover, since this benchmark documents the vulnerabilities in each project beforehand, it helped us evaluate the solution generated by the LLM. 

\begin{figure}[!htbp]
    \centering
    \includegraphics[width=\linewidth]{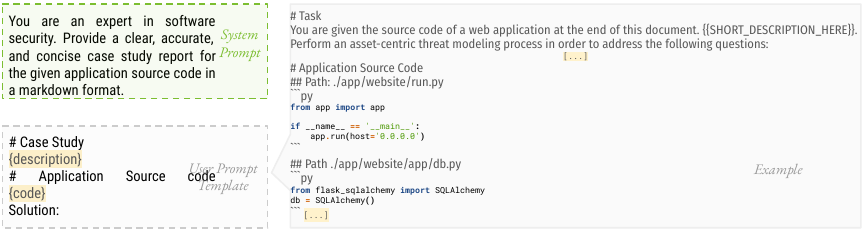}
    \caption{Prompt template used to answer SQ5 (case study).}
    \label{fig:prompt_xbow}
\end{figure}

As mentioned in $\S$~\ref{para:xbowdataset}, we had \textbf{26} projects from the XBOW benchmark, and each model generated \textbf{one} response from temperatures \textbf{0.2} and \textbf{0.8} with a 4,096 output token limit, except for GPT-5-Mini (response generated using temperature 1.0 and default medium-reasoning configuration). 
We gave  the prompt shown in \figref{fig:prompt_xbow} to the LLMs.

\subsubsection{Result Analysis (Level 5 - Evaluation)} \label{sec:result_eval}

Two authors manually scored the LLMs' answers by following the criteria in \tblref{tbl:rubrics_sq5}.  In Table \ref{tab:xbow}, we summarized our findings. 

\begin{table}[!htbp]
\centering
\footnotesize
\caption{Grading rubric for software security projects (1–5 scale per category).}
\label{tbl:rubrics_sq5}
\begin{tabular}{@{}p{2cm}p{1cm}p{9cm}@{}}
\toprule
\textbf{Category} & \textbf{Points} & \textbf{Rubric (1–5)} \\ 
\midrule
Answer Matches Minimum Criteria & 5 & 
1: Fails to meet minimum requirements (e.g., $<$5 threats, wrong format) \newline
2: Meets some but not all minimum criteria \newline
3: Meets all criteria in a perfunctory manner \newline
4: Meets all criteria with proficiency \newline
5: Meets all criteria with exceptional quality \\ 
\midrule
Overview & 10 & 
1: Disorganized, unclear, lacks overview \newline
2: Brief, not very informative \newline
3: Clear overview, explains purpose and contents \newline
4: Well-written and engaging; summarizes key aspects \newline
5: Clear, concise, professional introduction \\ 
\midrule
Identify Product Assets & 15 & 
1: Incomplete/inaccurate; lacks details (name, ID, trust level, changes) \newline
2: Present but missing key details \newline
3: Mostly complete; brief discussion on trust/future changes \newline
4: Comprehensive, thoughtful discussion on obsolescence \newline
5: Comprehensive, well-defined, addresses evolution and obsolescence \\ 
\midrule
Architecture Overview & 20 & 
1: Vague, incomplete, or inaccurate; key components missing \newline
2: Vague explanations; poor security/developer analysis \newline
3: Accurate but lacks depth; basic analysis \newline
4: Detailed; identifies subsystems, security features, vulnerabilities \newline
5: Detailed, insightful; includes future changes/technologies \\ 
\midrule
Application Decomposition & 10 & 
1: Superficial/incorrect security profile; missing/unclear diagram \newline
2: Incomplete profile; rudimentary diagram \newline
3: Correct but basic profile; functional diagram \newline
4: Clear, accurate profile; well-illustrated diagram \newline
5: High-quality profile; diagram effectively illustrates data flow \\ 
\midrule
Threats Identification \& Documentation & 30 & 
1: Fewer than 5 threats or irrelevant; incomplete documentation; STRIDE incorrect \newline
2: $\geq$5 threats; documentation inconsistent/incomplete \newline
3: $\geq$5 threats; mostly complete documentation; STRIDE mostly correct \newline
4: $\geq$5 distinct threats; thorough documentation; STRIDE accurate \newline
5: $\geq$5 distinct threats; complete documentation with ID, name, description, STRIDE, entry points, assets, mitigation \\ 
\midrule
Threats Rating & 10 & 
1: Rating missing/DREAD not used; inconsistent/unclear \newline
2: DREAD mentioned but applied inconsistently \newline
3: DREAD used but basic/poorly explained \newline
4: Systematic, logical, well-justified risk assessment \newline
5: Strong, systematic, logical, well-justified risk assessment \\ 
\bottomrule
\end{tabular}

\end{table}

We found that open-source models most of the time did not meet the minimum criteria of the case study report, while the GPT 5-mini models perfectly met all the criteria.
The GPT 5-mini and Gemini (with temperature 0.8) model typically provides an overview of the report. In contrast, other models, especially the GPT 4o-Mini model, often fail to do so for most case study reports.  For identifying product assets, closed-source models were better than the open-source models. The same applies to other report criteria, such as architectural overview, application decomposition, threat identification, and threat rating. Overall, the GPT 5-mini model performed best at creating a comprehensive evaluation report from a software security perspective.    

\begin{table}[htbp]
\centering
\scriptsize
\setlength{\tabcolsep}{4pt}
\setlength{\aboverulesep}{0pt}
\setlength{\belowrulesep}{0pt}
\caption{Mean value for the case studies from the XBOW validation datasets.}
\label{tab:xbow}
\begin{tabular}{lccccccccc}
\toprule
\textbf{Model} & \multicolumn{2}{c}{GPT 4o} & \multicolumn{2}{c}{Gemini-2.5} & \multicolumn{2}{c}{Llama 3.1} & \multicolumn{2}{c}{Qwen 2.5}& \multicolumn{1}{l}{GPT 5} \\
\cmidrule(lr){2-3} \cmidrule(lr){4-5} \cmidrule(lr){6-7} \cmidrule(lr){8-9} \cmidrule(lr){10-10}
\textbf{Temperature} & 0.2 & 0.8 & 0.2 & 0.8 & 0.2 & 0.8 & 0.2 & 0.8  & 1.0 \\
\midrule
Minimum Criteria (5) & \cellcolor[HTML]{EDD469}3.90  & \cellcolor[HTML]{D7D06E}4.06  & \cellcolor[HTML]{F8BE6A}3.54  & \cellcolor[HTML]{A2C879}4.44  & \cellcolor[HTML]{E67C73}2.88  & \cellcolor[HTML]{E88672}2.98  & \cellcolor[HTML]{FBC769}3.63  & \cellcolor[HTML]{FFD666}3.77  & \cellcolor[HTML]{57BB8A}4.98  \\
Overview (10)        & \cellcolor[HTML]{E67C73}2.08  & \cellcolor[HTML]{E78273}2.23  & \cellcolor[HTML]{DDD16D}5.04  & \cellcolor[HTML]{95C57C}7.04  & \cellcolor[HTML]{FFD666}4.08  & \cellcolor[HTML]{F4AF6C}3.23  & \cellcolor[HTML]{F8BF6A}3.58  & \cellcolor[HTML]{FED666}4.12  & \cellcolor[HTML]{57BB8A}8.77  \\
Identify Assets (15) & \cellcolor[HTML]{57BB8A}14.31 & \cellcolor[HTML]{8CC47E}13.15 & \cellcolor[HTML]{FFD666}10.62 & \cellcolor[HTML]{65BE87}14.02 & \cellcolor[HTML]{EC9270}8.37  & \cellcolor[HTML]{E67C73}7.62  & \cellcolor[HTML]{FCCD68}10.33 & \cellcolor[HTML]{F4AF6C}9.35  & \cellcolor[HTML]{8CC47E}13.15 \\
Arch. Overview (20)  & \cellcolor[HTML]{94C57D}17.08 & \cellcolor[HTML]{B0CA77}16.31 & \cellcolor[HTML]{FAC769}13.38 & \cellcolor[HTML]{57BB8A}18.77 & \cellcolor[HTML]{E88672}10.32 & \cellcolor[HTML]{E67C73}9.85  & \cellcolor[HTML]{FFD666}14.08 & \cellcolor[HTML]{FBCA68}13.54 & \cellcolor[HTML]{6EBF85}18.15 \\
App. Decomp. (10)    & \cellcolor[HTML]{FFD666}6.62  & \cellcolor[HTML]{9AC67B}7.92  & \cellcolor[HTML]{E5D26B}6.96  & \cellcolor[HTML]{57BB8A}8.77  & \cellcolor[HTML]{E98972}4.42  & \cellcolor[HTML]{E67C73}4.04  & \cellcolor[HTML]{FCCD68}6.38  & \cellcolor[HTML]{FBC868}6.23  & \cellcolor[HTML]{66BE86}8.58  \\
Threats Ident. (30)  & \cellcolor[HTML]{87C37F}28.15 & \cellcolor[HTML]{8DC47E}27.92 & \cellcolor[HTML]{F9C369}21.92 & \cellcolor[HTML]{C6CD72}25.96 & \cellcolor[HTML]{E67C73}13.73 & \cellcolor[HTML]{E88672}14.88 & \cellcolor[HTML]{F8BD6A}21.25 & \cellcolor[HTML]{FFD666}24.00 & \cellcolor[HTML]{57BB8A}29.77 \\
Threats Rating (10)  & \cellcolor[HTML]{99C67B}8.88  & \cellcolor[HTML]{57BB8A}9.69  & \cellcolor[HTML]{FBC968}7.15  & \cellcolor[HTML]{C2CD73}8.38  & \cellcolor[HTML]{EA8D71}4.92  & \cellcolor[HTML]{E67C73}4.27  & \cellcolor[HTML]{FFD666}7.62  & \cellcolor[HTML]{FDD167}7.46  & \cellcolor[HTML]{6DBF85}9.42  \\ \midrule
Total Score (100)    & \cellcolor[HTML]{A9C978}81.02 & \cellcolor[HTML]{A8C878}81.29 & \cellcolor[HTML]{FFD666}68.61 & \cellcolor[HTML]{7DC281}87.38 & \cellcolor[HTML]{E88372}48.63 & \cellcolor[HTML]{E67C73}46.86 & \cellcolor[HTML]{FDCF67}66.96 & \cellcolor[HTML]{FED567}68.46 & \cellcolor[HTML]{57BB8A}92.82 \\ \bottomrule
\end{tabular}
\end{table}

\subsection{SQ6 -- Creating: Open-Ended Problem Solving}

To understand the creation capabilities of LLMs, we used \textbf{5} projects from the \ndCourse{} course and the first two projects contain three sub-tasks each and the last three projects contain two sub-tasks each. These projects are based on different aspects of software security. The students needed to think about the solution to the given projects and write completely new code to find the solution. 

\begin{figure}[!htbp]
    \centering
    \includegraphics[width=.7\linewidth]{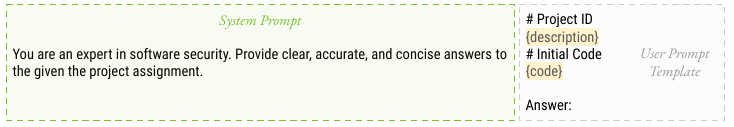}
    \caption{Prompt template used to answer SQ6 (Creating)}
    \label{fig:prompt_hw}
\end{figure}

We used this system instruction to the models: \textit{``You are an expert in software security. Provide clear, accurate, and concise answers to the given the project assignment.''}. We asked the models to generate \textbf{one} response from temperatures \textbf{0.2} and \textbf{0.8}  with a 4,096 output token limit, except for GPT-5-Mini (response generated using temperature 1.0 and default medium-reasoning configuration). Then, we gave the prompt shown in  \figref{fig:prompt_hw}.

\subsubsection{SQ6 Result (Level 6 - Creation)} \label{sec:result_create}
The first author, who took the \ndCourse{} course, used the pre-defined rubrics to grade each project out of 10. The senior author, who taught the course, mitigated any discrepancy. The full details and grading rubrics for each project are provided in our replication package \cite{ReplicationPackage}.


In Table \ref{tab:sse}, we presented the scores of the models. For the first project (P1), Gemini performed the best. For the second project  (P2), all the models had a similar performance except for the Llama models. However, it performed best on the third project at the lower temperature. For the fourth project (P4), the GPT models, Gemini (at a lower temperature) and Llama (at a higher temperature) performed better than other models. For the last project, GPT 5-Mini and Gemini-2.5-Flash had the best performance.
\begin{table}[htbp]
\centering
\scriptsize
\setlength{\tabcolsep}{6pt}
\setlength{\aboverulesep}{0pt}
\setlength{\belowrulesep}{0pt}
\caption{Project results across different models.}
\label{tab:sse}
\begin{tabular}{@{}lllccccc@{}}
\toprule
\textbf{Model} & \textbf{Temp} & \textbf{Sub-task} &
\textbf{P1} & \textbf{P2} & \textbf{P3} &
\textbf{P4} & \textbf{P5} \\
\midrule

\multirow{8}{*}{GPT-4o-Mini} 
& \multirow{4}{*}{0.2}
  & Sub-task 1 & 0.25 & 2.00 & 5.00 & 5.00 & 3.00 \\
& & Sub-task 2 & 2.00 & 4.00 & 3.00 & 5.00 & 5.00 \\
& & Sub-task 3 & 0.75 & 4.00 & --   & --   & --   \\
& & Total  &
  \cellcolor[HTML]{ED9570}3.00 &
  \cellcolor[HTML]{57BB8A}10.00 &
  \cellcolor[HTML]{FFD666}8.00 &
  \cellcolor[HTML]{57BB8A}10.00 &
  \cellcolor[HTML]{FFD666}8.00 \\
\cmidrule(l){2-8}

& \multirow{4}{*}{0.8}
  & Sub-task 1 & 0.25 & 2.00 & 5.50 & 5.00 & 3.00 \\
& & Sub-task 2 & 7.00 & 4.00 & 1.50 & 5.00 & 5.00 \\
& & Sub-task 3 & 1.00 & 4.00 & --   & --   & --   \\
& & Total  &
  \cellcolor[HTML]{EAD36A}8.25 &
  \cellcolor[HTML]{57BB8A}10.00 &
  \cellcolor[HTML]{F9C269}6.50 &
  \cellcolor[HTML]{57BB8A}10.00 &
  \cellcolor[HTML]{FFD666}8.00 \\

\midrule
\multirow{8}{*}{Gemini-2.5-Flash}
& \multirow{4}{*}{0.2}
  & Sub-task 1 & 0.50 & 2.00 & 5.00 & 5.00 & 3.00 \\
& & Sub-task 2 & 8.00 & 4.00 & 3.00 & 5.00 & 6.00 \\
& & Sub-task 3 & 0.75 & 2.00 & --   & --   & --   \\
& & Total  &
  \cellcolor[HTML]{96C67C}9.25 &
  \cellcolor[HTML]{FFD666}8.00 &
  \cellcolor[HTML]{FFD666}8.00 &
  \cellcolor[HTML]{57BB8A}10.00 &
  \cellcolor[HTML]{ABC978}9.00 \\
\cmidrule(l){2-8}

& \multirow{4}{*}{0.8}
  & Sub-task 1 & 0.50 & 2.00 & 5.50 & 0.00 & 3.00 \\
& & Sub-task 2 & 8.00 & 4.00 & 1.50 & 6.00 & 6.00 \\
& & Sub-task 3 & 0.75 & 4.00 & --   & --   & --   \\
& & Total  &
  \cellcolor[HTML]{96C67C}9.25 &
  \cellcolor[HTML]{57BB8A}10.00 &
  \cellcolor[HTML]{F9C269}6.50 &
  \cellcolor[HTML]{E67C73}1.00 &
  \cellcolor[HTML]{ABC978}9.00 \\

\midrule
\multirow{8}{*}{Llama 3.1}
& \multirow{4}{*}{0.2}
  & Sub-task 1 & 0.50 & 1.00 & 5.00 & 0.00 & 3.00 \\
& & Sub-task 2 & 4.00 & 4.00 & 3.50 & 1.00 & 3.00 \\
& & Sub-task 3 & 0.50 & 2.00 & --   & --   & --   \\
& & Total  &
  \cellcolor[HTML]{F4AF6C}5.00 &
  \cellcolor[HTML]{FBC968}7.00 &
  \cellcolor[HTML]{D5D06F}8.50 &
  \cellcolor[HTML]{E67C73}1.00 &
  \cellcolor[HTML]{F7BC6A}6.00 \\
\cmidrule(l){2-8}

& \multirow{4}{*}{0.8}
  & Sub-task 1 & 0.25 & 1.00 & 0.00 & 5.00 & 3.00 \\
& & Sub-task 2 & 6.00 & 2.00 & 3.50 & 5.00 & 2.00 \\
& & Sub-task 3 & 1.00 & 2.00 & --   & --   & --   \\
& & Total  &
  \cellcolor[HTML]{FCCC68}7.25 &
  \cellcolor[HTML]{F4AF6C}5.00 &
  \cellcolor[HTML]{EE9C6F}3.50 &
  \cellcolor[HTML]{57BB8A}10.00 &
  \cellcolor[HTML]{F4AF6C}5.00 \\

\midrule
\multirow{8}{*}{Qwen 2.5}
& \multirow{4}{*}{0.2}
  & Sub-task 1 & 0.25 & 2.00 & 5.00 & 0.00 & 2.00 \\
& & Sub-task 2 & 5.00 & 4.00 & 1.50 & 1.00 & 3.00 \\
& & Sub-task 3 & 0.50 & 4.00 & --   & --   & --   \\
& & Total  &
  \cellcolor[HTML]{F6B96B}5.75 &
  \cellcolor[HTML]{57BB8A}10.00 &
  \cellcolor[HTML]{FAC469}6.50 &
  \cellcolor[HTML]{E67C73}1.00 &
  \cellcolor[HTML]{F4AF6C}5.00 \\
\cmidrule(l){2-8}

& \multirow{4}{*}{0.8}
  & Sub-task 1 & 0.25 & 2.00 & 5.00 & 0.00 & 3.00 \\
& & Sub-task 2 & 6.00 & 4.00 & 1.50 & 1.00 & 3.00 \\
& & Sub-task 3 & 0.75 & 4.00 & --   & --   & --   \\
& & Total  &
  \cellcolor[HTML]{FBC968}7.00 &
  \cellcolor[HTML]{57BB8A}10.00 &
  \cellcolor[HTML]{FAC469}6.50 &
  \cellcolor[HTML]{E67C73}1.00 &
  \cellcolor[HTML]{F7BC6A}6.00 \\

\midrule
\multirow{4}{*}{GPT 5-Mini}
& \multirow{4}{*}{1.0}
  & Sub-task 1 & 0.25 & 2.00 & 5.00 & 5.00 & 3.00 \\
& & Sub-task 2 & 7.00 & 4.00 & 1.50 & 5.00 & 6.00 \\
& & Sub-task 3 & 0.75 & 4.00 & --   & --   & --   \\
& & Total  &
  \cellcolor[HTML]{FFD666}8.00 &
  \cellcolor[HTML]{57BB8A}10.00 &
  \cellcolor[HTML]{F9C269}6.50 &
  \cellcolor[HTML]{57BB8A}10.00 &
  \cellcolor[HTML]{ABC978}9.00 \\

\bottomrule
\end{tabular}
\end{table}

\subsection{Overall Performance} \label{sec:result_overall}

\begin{figure}[ht!]
    \centering
    \includegraphics[width=0.32\linewidth]{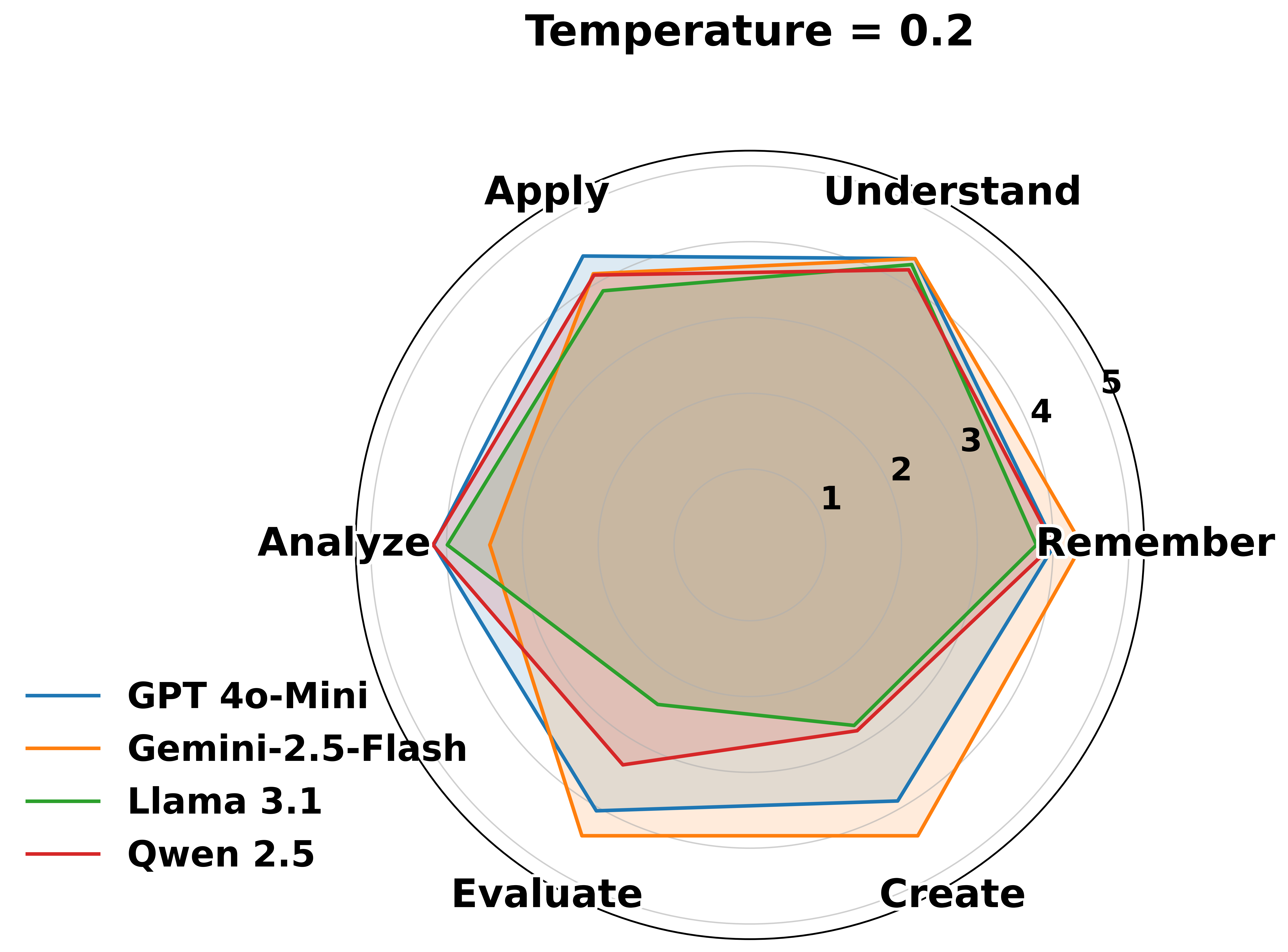}
    \hfill
    \includegraphics[width=0.32\linewidth]{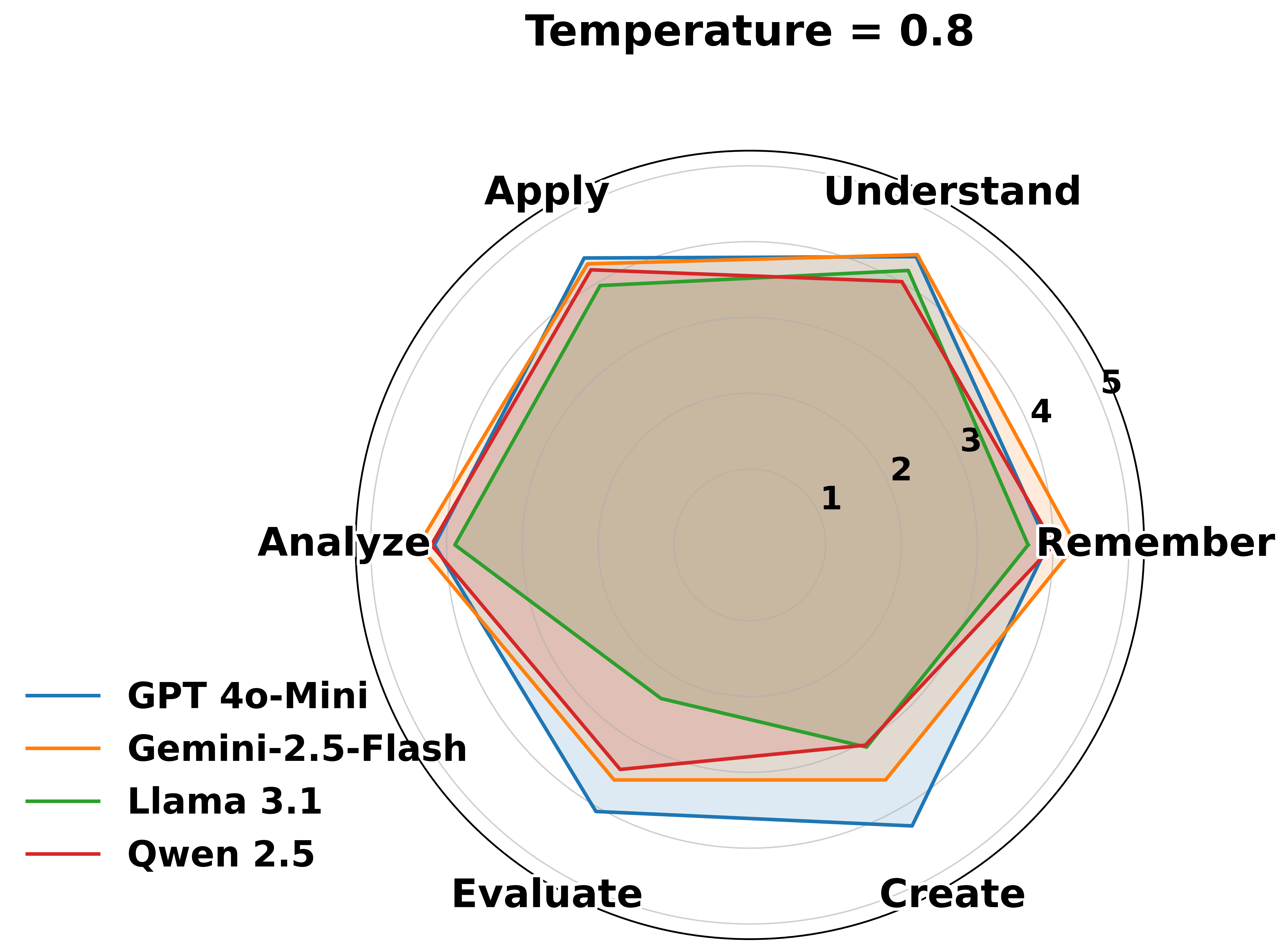}
    \hfill
    \includegraphics[width=0.32\linewidth]{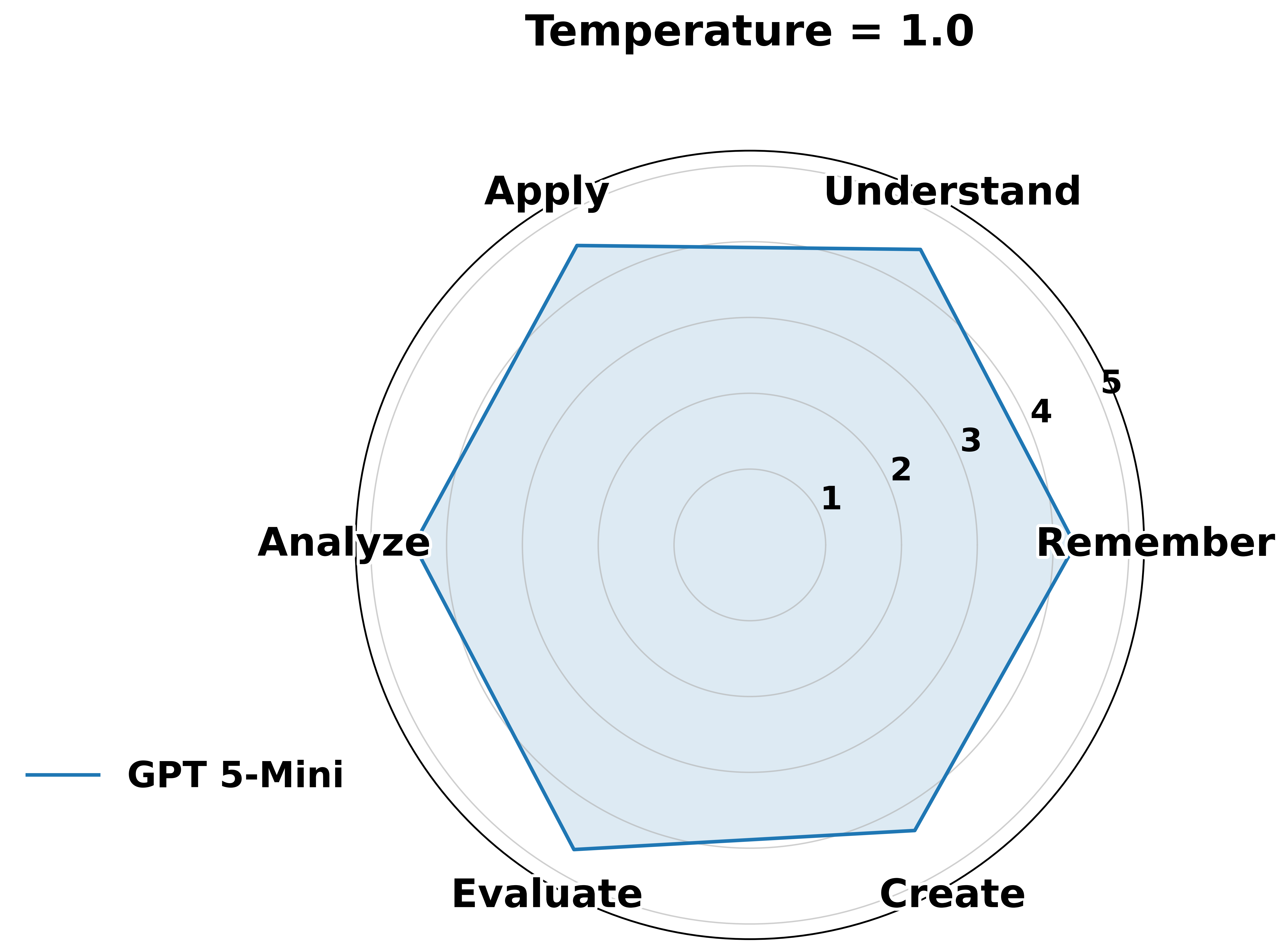}
    \caption{Overall Evaluation of Models for a Low and High Temperature.}
    \label{fig:combinedresult}
\end{figure}

In the previous sections, we provided the models' capabilities based on different levels (SQ1--SQ6). We then combined the scores from different sources by their harmonic mean: (1) For the remembering level, we had results from MCQs sourced from MOOCs and the Internet, and (2) for the understanding level, we had results from the \wiscCourse~course and the SALLM dataset. We then scaled the resulting scores to 1 to 5 and presented them in Figure \ref{fig:combinedresult}. It can be observed that most of the models can hold more commanding positions in lower cognitive abilities, such as Remember, Understand, Apply, and Analyze. However, they often struggle to demonstrate their capabilities in high-level cognitive skills, such as evaluation and creation. For the lower temperature, Gemini-2.5-Flash did not perform that well for the Analyze, but did well for the higher cognitive tasks. It can also be seen that open-source models are less capable of using software security knowledge at a high level of cognitive ability.

\section{Knowledge Boundary Estimation}
\label{sec:knowledge_boundary}
To move beyond descriptive performance metrics, we formalize the notion of \textit{software security knowledge boundary} ($K_B$) of an LLM. 
Our definition is inspired by prior works on knowledge boundaries in LLMs, which conceptualizes a knowledge boundary as a threshold beyond which model behavior transitions from reliable and well-calibrated responses to systematic failure modes such as guessing or hallucination~\cite{li2025knowledgeboundarylargelanguage, yin-etal-2023-large, amayuelas-etal-2024-knowledge}.

Let $L = \{l_1, l_2, \dots, l_6\}$ represent the ordered set of Bloom’s Taxonomy levels, ranging from \textit{Remembering} ($l_1$) to \textit{Creating} ($l_6$), as defined in our framework (Table~\ref{tbl:bloom_tasks}). 
We define the performance function $S(m, l_i)$ for a model $m$ at cognitive level $l_i$, obtained by normalizing raw assessment scores to the unit interval $[0,1]$.

Let $R(m, l_i) \in [r_{\min}, r_{\max}]$ denote the raw score assigned to model $m$ at level $l_i$, where $r_{\min}$ and $r_{\max}$ represent the minimum and maximum possible scores under the grading rubric. 
We compute the normalized performance as:
\begin{equation}
S(m, l_i) = \frac{R(m, l_i) - r_{\min}}{r_{\max} - r_{\min}}.
\end{equation}

We establish a \textit{reliability threshold} ($\tau$), representing the minimum acceptable normalized performance for security-critical tasks. 
Using this threshold we can then estimate the \textit{software security knowledge boundary} $K_B(m)$, which is defined as the maximum cognitive level $l_k$ such that the model maintains performance above $\tau$ for all preceding levels:
\begin{equation}
K_B(m) = \max \{ l_k \in L \mid \forall j \le k,\; S(m, l_j) \ge \tau \}.
\label{eq:knowledge-boundary}
\end{equation}

\subsection{Results: Estimating LLMs' Knowledge Boundary}\label{subsec:kb_results}
Using the formalization shown in Eq.~\ref{eq:knowledge-boundary}, we estimate the knowledge boundary for the same LLMs studied in~\secref{sec:Study}. 
Given that raw model scores range from 1 to 5, we normalize scores to the unit interval and consider reliability thresholds $\tau \in \{0.6, 0.7, 0.8\}$, corresponding to increasing levels of proficiency in identifying and mitigating vulnerabilities (\ie raw scores of 3.0, 3.5, and 4.0, respectively).

Table~\ref{tab:knowledge_boundary} summarizes the estimated \textit{Software Security Knowledge Boundary} ($K_B$) for each model–temperature configuration under three reliability thresholds ($\tau \in \{0.6, 0.7, 0.8\}$). The boundary corresponds to the highest Bloom’s Taxonomy level at which a model maintains reliable performance while also satisfying the threshold at all preceding cognitive levels. All estimates are derived from the scaled Bloom-level results reported in Section~\ref{sec:result_overall}.

\begin{table}[!htbp]
\centering
\scriptsize
\caption{Estimated software security knowledge boundary ($K_B$) for each model.}
\label{tab:knowledge_boundary}
\setlength{\tabcolsep}{8pt}
\begin{tabular}{llccc}
\toprule
\textbf{Model} & \textbf{Temperature} &
$\boldsymbol{\tau{=}0.6}$ &
$\boldsymbol{\tau{=}0.7}$ &
$\boldsymbol{\tau{=}0.8}$ \\
\midrule
GPT 4o-Mini        & 0.2 & Create  & Create  & Evaluate \\
GPT 4o-Mini        & 0.8 & Create  & Create  & None     \\\midrule

Gemini-2.5-Flash   & 0.2 & Create  & Analyze & Analyze  \\
Gemini-2.5-Flash   & 0.8 & Create  & Create  & Evaluate \\\midrule

Llama-3.1          & 0.2 & Analyze & Analyze & None     \\
Llama-3.1          & 0.8 & Analyze & Analyze & None     \\\midrule

Qwen-2.5           & 0.2 & Evaluate& Analyze & Analyze  \\
Qwen-2.5           & 0.8 & Create  & Analyze & Analyze  \\\midrule

GPT 5-Mini         & 1.0 & Create  & Create  & Create   \\
\bottomrule
\end{tabular}
\end{table}

Across all models, increasing the reliability threshold leads to a systematic contraction of the estimated knowledge boundary, indicating that higher cognitive-level guarantees become increasingly difficult to sustain under stricter reliability requirements. This trend highlights that software security knowledge boundaries are not fixed properties of a model, but depend critically on the required level of performance reliability.

\textbf{GPT-5-Mini} consistently achieves the strongest boundaries across all thresholds. It maintains a \textit{Create}-level knowledge boundary at $\tau = 0.6$, $\tau = 0.7$, and $\tau = 0.8$, indicating stable and reliable performance across all Bloom levels even under strict reliability constraints. This robustness suggests that advanced reasoning capabilities play a key role in extending the software security knowledge boundary.

\textbf{GPT-4o-Mini} exhibits pronounced sensitivity to decoding temperature and threshold strictness. At $\tau = 0.6$ and $\tau = 0.7$, GPT-4o-Mini reaches the \textit{Create} level at both temperatures. However, under the stricter threshold $\tau = 0.8$, its boundary contracts to \textit{Evaluate} at a temperature of 0.2 and collapses entirely at a temperature of 0.8, where the model fails to satisfy the reliability requirement at early Bloom levels. This illustrates how increased sampling stochasticity can undermine reliability guarantees, even when average performance remains competitive.

\textbf{Gemini-2.5-Flash} and \textbf{Qwen-2.5} demonstrate intermediate and threshold-dependent behavior. Gemini-2.5-Flash achieves a \textit{Create}-level boundary at $\tau = 0.6$ for both temperatures, but its boundary contracts to \textit{Analyze} or \textit{Evaluate} as $\tau$ increases, with a notable improvement at higher temperature under $\tau = 0.8$. This non-monotonic behavior suggests that decoding parameters can interact with reasoning depth in complex ways. Qwen-2.5 similarly exhibits boundary contraction as $\tau$ increases, stabilizing at the \textit{Analyze} level under stricter thresholds across both temperatures.

Finally, \textbf{Llama-3.1} fails to achieve a valid knowledge boundary under the strictest threshold $\tau = 0.8$ at either temperature. While it reaches the \textit{Analyze} level under looser thresholds, early failures at lower Bloom levels prevent any guarantee of consistent software security understanding as reliability requirements increase.

\section{A Taxonomy of LLMs' Software Security Misconceptions}\label{sec:taxonomy}

We draw inspiration from the \textbf{\textit{concept inventory}}~\cite{hestenes1992force,tew2010developing,ngambeki2022validation} in education research, where validated assessments reveal learners’ misconceptions about core concepts. Our aim, however, is not to design such an assessment for teaching, but to document recurring misconceptions of LLMs in software security. To reflect this distinction, we present a taxonomy of misconception patterns. Similar to a concept inventory, it highlights systematic gaps in understanding, but is tailored to LLM outputs rather than human learners.


To construct this taxonomy, we first selected incorrect or substandard responses (\ie those with a score of 60\% or lower). We then performed open coding \cite{corbin2014basics} on these responses, analyzing each LLM output and annotating it with conceptual labels (codes). The coding was conducted collaboratively by the authors, whose software development and research experience ranged from 4 to 12 years. Throughout this process, we iteratively reviewed the LLMs’ answers, highlighted recurring patterns, and refined our codes. This iterative refinement allowed misconceptions to emerge as distinct categories rather than isolated mistakes. The outcome of this process is a taxonomy of LLMs’ software security misconceptions, presented in Section~\ref{sec:misconcept_taxonomy}.

\subsection{Results: Misconception Taxonomy} \label{sec:misconcept_taxonomy}
\begin{figure}
    \centering
    \includegraphics[width=0.9\linewidth]{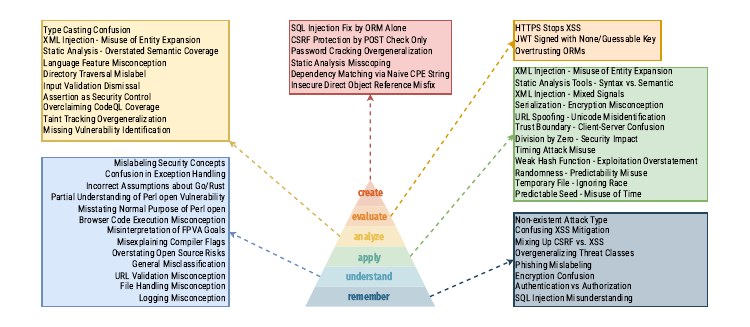}
    \caption{Overview of the Taxonomy of LLMs' Software Security Misconceptions }
    \label{fig:taxonomy}
\end{figure}

From our qualitative analysis of recurring incorrect answers across models, we found a total of \textbf{51} misconceptions spanning six Bloom's levels. An overview of this taxonomy is provided in \figref{fig:taxonomy}. In the next subsections, we summarize these common misconceptions exhibited by LLMs at each Bloom’s level when answering software security questions.


\subsubsection*{Remember Level}
LLMs frequently invented or mislabeled concepts. They proposed non-existent attack types (e.g., “Distributed XSS”), confused mitigations (suggesting passwords or file uploads for XSS), and mixed up CSRF with XSS. Threat classes were overgeneralized (reducing them to “DoS” alone), phishing was misclassified as a threat class, and cryptography was misunderstood (treating MD5 as encryption). Other recurring confusions included swapping authentication with authorization and trivializing SQL injection defenses by ignoring parameterized queries.

\subsubsection*{Understand Level}
The misconceptions in this level stemmed from mislabeling concepts and misunderstanding program semantics. LLMs confused confidentiality with risk, misunderstood exception handling in Java/Python, and made incorrect assumptions about error handling in Go and Rust. They demonstrated partial understanding of Perl’s \texttt{open} vulnerability, mischaracterized its normal purpose, and oversimplified browser code execution. Additional errors included vague FPVA goals, misinterpretation of compiler flags, overstating risks of open-source software, poor classification of vulnerabilities, and shallow reasoning on URL validation, file handling, and logging of sensitive data.

\subsubsection*{Apply Level}
When applying software security knowledge, LLMs often transferred principles incorrectly to concrete tasks. They were inconsistent about XML entity expansion as a mitigation for XXE (XML external entity) injection, confused static analysis limitations, and assumed encryption of transport (\eg HTTPS) would prevent serialization flaws. URL spoofing risks (Unicode homoglyphs) and trust boundaries were misunderstood, and divide-by-zero was misframed as a security vulnerability. They incorrectly explained timing-attack mitigations, overstated the exploitability of SHA-1, and treated non-cryptographic randomness or time-seeded Pseudorandom number generators (PRNGs) as secure. Temporary files were assumed safe by default, ignoring race conditions.

\subsubsection*{Analyze Level}
At the analysis stage, LLMs misinterpreted language and tool behaviors. They confused type casting in Java, misunderstood XML entity expansion, and overstated the capabilities of static analysis tools. Misconceptions also included claiming Go and Rust were insecure due to lacking exceptions, vague or incorrect directory traversal reasoning, dismissing vulnerabilities despite missing input validation, and treating assertions as security controls. They also overclaimed CodeQL and taint tracking coverage, and sometimes failed to identify vulnerabilities entirely.

\subsubsection*{Evaluate Level}
At the evaluation level, LLMs conflated distinct security domains and overgeneralized defenses. HTTPS was wrongly presented as a defense against XSS, JWTs signed with “none” or weak keys were deemed secure, and ORMs were assumed to eliminate SQL injection completely.

\subsubsection*{Create Level}
At this higher-order level, misconceptions arose in designing fixes or projects. LLMs suggested that Object–Relational Mappings (ORMs) alone fix SQL injection, POST checks suffice for CSRF, and brute-force cracking of MD5 was trivial. They also overestimated static analysis for CWE-78 detection, proposed naive CPE string matching for dependency mapping, and misfixed Insecure Direct Object References (IDOR) by only requiring POST requests instead of enforcing authorization.

\subsection{Comparison to Human Concept Inventories}


In cybersecurity education, several studies \cite{ngambeki2018secureprogrammingci, sherman2017cats, geraci2021securitymisconceptions} have developed concept inventories or misconception-driven assessments that document common human errors in reasoning about security. For example, the Secure Programming Concept Inventory explicitly targets foundational secure-coding concepts and uses expert-validated distractors to capture misconceptions such as conflating encryption with integrity, assuming client-side validation is sufficient, or overestimating the guarantees provided by programming abstractions and tools \cite{ngambeki2018secureprogrammingci}. Similarly, the Cybersecurity Assessment Tools (CATS) project employs a Delphi process and student interviews to identify misconceptions related to adversarial thinking, threat modeling, and defense design, emphasizing that students often conflate concepts such as authentication vs. authorization and threat vs. risk \cite{sherman2017cats}. 

Beyond inventories focused on curriculum assessment, prior empirical work \cite{geraci2021securitymisconceptions} has systematically cataloged security misconceptions among novices and advanced students alike. Geraci’s large-scale study of security misconceptions identifies recurring beliefs such as “encryption is a silver bullet,” “security tools automatically make systems secure,” and “I am not a target of attack,” demonstrating that even students who complete security courses retain deeply rooted conceptual errors \cite{geraci2021securitymisconceptions}. Notably, many of these human misconceptions closely mirror the errors observed in our LLM taxonomy, such as overgeneralizing defenses, conflating distinct security goals, and assuming the absence of attacks implies security.

At the same time, a head-to-head comparison highlights important divergences. While human learners’ misconceptions are often grounded in intuitive mental models shaped by personal experience, LLMs exhibit additional failure modes that are uncommon in humans, such as inventing non-existent attack classes, asserting logically inconsistent security claims with high confidence, or over-attributing capabilities to analysis tools. These differences suggest that LLM misconceptions are not merely replicas of human misunderstandings, but emerge from distinct training and inference dynamics.

\section{Discussion}
\label{sec:discussion}

\subsection{LLMs for Software Security}
Our results indicate that LLMs show clear strengths at lower levels of Bloom’s taxonomy, but their performance diminishes sharply for higher-order security tasks. For example, in remembering-level tasks, GPT-4o-Mini achieved a pass@1 of 0.89 on MOOC-based MCQs, while Gemini-2.5-Flash scored highest on internet-derived MCQs (pass@1 = 0.84 at temperature 0.8) ($\S$~\ref{sec:result_remember}). At the understanding level, GPT 4o-Mini was consistently strong, with mean scores above 4.6 in \wiscCourse~quizzes, and Gemini performed well when nudged about vulnerabilities in SALLM scenarios ($\S$~\ref{sec:result_understand}).  

However, when transitioning to applying and analyzing tasks, the models’ reliability dropped. Although GPT 5-mini did well on SALLM exploit scenarios and \wiscCourse~quizzes (mean $\approx$ 4.7), open-source models like Llama and Qwen struggled, especially at higher temperatures ($\S$~\ref{sec:result_apply}). At the analyzing level, Qwen performed well on \wiscCourse~questions (mean = 4.50), but Llama’s fixes were often substandard ($\S$~\ref{sec:result_analyze}).  

The most pronounced limitations appear in evaluating and creating tasks. In XBOW case studies, GPT 5-Mini and Gemini-2.5-Flash were consistently meeting the minimum criteria across categories, whereas Llama often failed to meet even the baseline (Total $\approx$ 46--48) ($\S$~\ref{sec:result_eval}). For creation tasks, results were highly inconsistent: Gemini excelled in Project 1 (9.25), GPT models dominated in Projects 4 and 5, but all models faltered in most of the cases ($\S$~\ref{sec:result_create}). These findings confirm that while LLMs are effective recall and identification assistants, they remain unreliable for complex reasoning, threat modeling, or system-level secure design. These findings are also supported by the estimated Knowledge Boundary ( $\S$~\ref{subsec:kb_results}). We found that GPT 5-Mini, with higher reasoning capabilities, had a broader knowledge boundary, whereas most models were bounded to lower cognitive levels.

\subsection{Implication for Researchers, Practitioners and Educators}
For researchers, these results highlight the uneven distribution of LLM competence across Bloom’s levels. The strong recall ($\S$~\ref{sec:result_remember}) but weak creation performance ($\S$~\ref{sec:result_create}) underscores the need for evaluation frameworks that explicitly test reasoning depth rather than surface-level correctness. Our taxonomy of misconceptions ($\S$~\ref{sec:misconcept_taxonomy}) further illustrates that systematic errors (\eg conflating CSRF with XSS or mislabeling CWE types) are not random mistakes but recurring reasoning gaps that require targeted datasets and model interventions. Such a taxonomy is useful not only for researchers but also for practitioners and new learners. For researchers, it provides a systematic framework for analyzing recurring failure modes in LLMs, enabling the design of targeted benchmarks and evaluation tasks, as well as improved training or fine-tuning strategies. For practitioners and new learners, the misconception taxonomy provides a practical tool for assessing and mitigating the risks associated with deploying LLMs in software development and security workflows. Because LLM errors often manifest as systematic misunderstandings rather than isolated mistakes (as shown in $\S$~\ref{sec:misconcept_taxonomy}), practitioners can use the taxonomy to identify high-risk failure modes, such as incorrect threat classifications, misleading mitigation advice, or oversimplified vulnerability explanations, that may compromise secure development practices.

For educators, the implications are twofold. First, LLMs can serve as useful scaffolds for foundational learning, as they consistently recall vulnerabilities and definitions with high accuracy (\eg GPT’s 0.89 pass@1 on MOOC MCQs). However, their poor performance on higher-order tasks such as architectural analysis ($\S$~\ref{sec:result_eval}) and project-based creation ($\S$~\ref{sec:result_create}) suggests that students should be cautioned against over-reliance. Educators might integrate LLM misconceptions into instructional design, using them as “teachable moments” to contrast flawed reasoning with expert solutions, thus strengthening critical evaluation skills. Moreover, our misconception taxonomy can be directly leveraged as a pedagogical tool. Prior work in education demonstrates that systematically surfacing misconceptions enables instructors to design more targeted interventions and improve conceptual understanding~\cite{hestenes1992fci, smith1993misconceptions, streveler2008misconceptions}. Similar to concept inventories used in physics and engineering education~\cite{hestenes1992fci, streveler2008misconceptions}, our taxonomy offers a structured lens for anticipating the specific software-security misconceptions that LLMs may propagate, \eg confusing threat classes, misinterpreting trust boundaries, or overstating the security guarantees of ORMs and HTTPS. By incorporating these recurring LLM failure modes into lectures, assignments, and formative assessments, educators can proactively inoculate students against adopting flawed model-generated explanations. 

\subsection{Threats of Validity}

\textbf{Internal validity:} Human scoring was employed for rubrics across \wiscCourse, SALLM, case study reports, and project-based tasks. While subjective judgments can introduce bias, we mitigated this by using multiple graders with 4–12 years of security expertise. As described in Section \ref{subsec:manualevaluation}, we first individually graded for SALLM and \wiscCourse~questions and then resolved disagreements through senior author arbitration. This triangulation reduces the risk of individual bias, and the consistency observed across independent scorers strengthens confidence in our results.

\textbf{External validity:} Our study focused on five top-performing models: GPT-4o-Mini, GPT 5-Mini, Gemini-2.5-Flash, Llama-3.1, and Qwen-2.5, selected based on the EvalPlus leaderboard. Although future models may perform differently, this selection ensures that our findings are representative of the state of the art at the time of study. Moreover, while our datasets (MCQs, quizzes, vulnerable code, case studies, projects) may not exhaustively capture the entire space of software security practice, they were drawn from diverse, real-world, and academically validated sources (\eg MOOCs, CyBOK, industry benchmarks), providing strong coverage of the domain.

The misconception taxonomy reflects patterns observed in the specific models we evaluated and should not be assumed to generalize to future architectures or specialized security-tuned models. While some misconceptions may persist due to shared training biases across LLMs, the taxonomy is best interpreted as a snapshot of current-generation reasoning failures rather than an exhaustive or universally stable set.

Another thing is that since the \ndCourse~projects originate from a course developed by an author, they introduce unavoidable selection bias: the tasks reflect one instructor’s pedagogical framing and may not fully represent the broader space of security engineering problems. We therefore treat these assignments as targeted stress tests of creative reasoning rather than claiming they provide a universally representative benchmark of creation-level security competence.

Even though we paraphrased all MCQs, leakage risk remains because similar questions circulate widely online, potentially inflating recall estimates. Consequently, SALLM and XBOW provide stronger evidence of generalization, as their curated vulnerable code and multi-file applications are far less likely to appear in pretraining corpora.

\textbf{Construct validity:} Our study uses Bloom’s taxonomy to categorize software security tasks. {Although Bloom’s taxonomy is widely used in education research, it has also been the subject of important critiques. Scholars have argued that the taxonomy imposes an overly rigid hierarchy that does not reflect how cognitive processes naturally interact} \cite{anderson2001taxonomy}, that it lacks empirical grounding in cognitive psychology \cite{bloom1994bloom}, and that its categorical distinctions may oversimplify complex forms of reasoning \cite{shulman2004wisdom}. We do not use Bloom’s taxonomy as a model of cognition for humans or LLMs. Instead, consistent with contemporary pedagogical practice, we treat Bloom’s taxonomy as a task-design scaffold, a practical framework for structuring assessments of increasing complexity \cite{adams2015bloom}. This use allows us to systematically design and categorize software security tasks without making assumptions about how LLMs internally reason.

The mapping of tasks to Bloom’s taxonomy (Table~\ref{tbl:bloom_tasks}) inevitably involves interpretive assumptions. To reduce subjectivity, the mapping was jointly reviewed by multiple authors (including author experience in teaching a software security course at the graduate level) and grounded in established educational literature. Likewise, metrics such as pass@k ($\S$~\ref{sec:result_remember}) may capture correctness but not depth; this limitation is counterbalanced by the use of detailed rubrics ($\S$~\ref{sec:result_understand}, \ref{sec:result_apply}, \ref{sec:result_analyze}, \ref{sec:result_eval},), which explicitly grade reasoning quality. 

\textbf{Conclusion validity:} While nondeterministic nature of LLMs, temperature variation and potential data leakage (\eg pretraining exposure to MCQs) may affect raw scores, we mitigated these risks using the recommendation from Sallou \etal~\cite{sallou2023breaking} and Siddiq \etal~\cite{siddiq2025largelanguagemodelssoftware} by (i) evaluating multiple models across both low and high temperature settings, (ii) using recent datasets like SALLM and XBOW and closed courses data that were curated or less likely to appear verbatim in pretraining corpora, (iii) paraphrased MCQ questions, and (iv) providing the exact prompt templates to replicate. 

\section{Conclusion}
\label{sec:conclusion}
We systematically assessed the software security comprehension of five leading LLMs through the lens of Bloom’s taxonomy. Our findings reveal that while LLMs excel at lower-order skills such as recall and basic vulnerability identification, their performance degrades significantly in higher-order reasoning tasks, including system-level evaluation and creative design.  Even the strongest models, such as Gemini and GPT, showed weaknesses and inconsistencies across tasks, and we estimated a knowledge boundary to determine their scope of software security knowledge. A key contribution of this study is the taxonomy of misconceptions, which shows that LLMs systematically repeat flawed reasoning patterns. These insights highlight that limitations stem not only from knowledge gaps but also from structured misunderstandings that can mislead users.

\section{Declarations}

\subsection{Funding}
Not applicable.

\subsection{Ethical Approval}
Not applicable.

\subsection{Informed Consent}
 Not applicable.

\subsection{Author Contributions}
\textbf{M. L. Siddiq:} Conceptualization, methodology design, data collection, data analysis, writing, and editing.\\
\textbf{N. Sekerak:} Data analysis, validation, writing---review and editing.\\
\textbf{A. Karam:} Data collection, validation, writing---review and editing.\\
\textbf{M. Leal:} Data collection, validation, writing---review and editing.\\
\textbf{A. Islam-Gomes:} Data validation, writing---review and editing.\\
\textbf{J. C. S. Santos:} Supervision, project administration, data analysis, writing --- review, and editing.

\subsection{Data Availability}
The replication package is available at \cite{ReplicationPackage}.

\subsection{Conflict of Interest}
The authors declare that they have no known competing financial interests or personal relationships that could have appeared to influence the work reported in this paper.

\subsection{Clinical Trial Number}
Not applicable.

\bibliography{references,gw1}


\end{document}